\documentclass[a4paper,11pt]{article}
\usepackage{amssymb}
\usepackage{amsfonts}
\usepackage{afterpage}
\usepackage{dsfont}
\usepackage{bbm}
\usepackage{booktabs}
\usepackage{amsmath}
\usepackage{amsthm}
\usepackage{url}
\usepackage{pdflscape}
\usepackage{array}
\usepackage{arydshln}
\usepackage{verbatim}
\usepackage{pifont}
\usepackage{eurosym}

\usepackage[authoryear]{natbib}
\usepackage{rotating} 
\usepackage{tablefootnote}

\usepackage{multirow}							
\usepackage{arydshln}							
\usepackage{dsfont}
\usepackage{subcaption}
\usepackage{caption}
\usepackage{float}
\usepackage{algorithm,algcompatible}
\usepackage{graphicx}

\newcommand{\LineComment}[1]{\Statex \hfill\textit{#1}}
\usepackage[section]{placeins} 

\makeatletter
\AtBeginDocument{%
  \expandafter\renewcommand\expandafter\subsection\expandafter{%
    \expandafter\@fb@secFB\subsection
  }%
}
\makeatother

\algnewcommand\INPUT{\item[\textbf{Input:}]}%
\algnewcommand\OUTPUT{\item[\textbf{Output:}]}%

\DeclareMathOperator*{\sign}{sgn}

\makeatletter
\let\@fnsymbol\@arabic
\makeatother

\makeatletter
\renewcommand{\ALG@name}{Method}
\makeatother

\begin{document}
	
	\renewcommand{\baselinestretch}{1.4}\normalsize
	
	\title{\vspace{-2.0cm} Robust Knockoffs for Controlling False Discoveries With an Application to Bond Recovery Rates
	}
	
	\author{Konstantin G\"{o}rgen \thanks{Institute of Economics (ECON), Karlsruhe Institute of Technology, Bl\"{u}cherstr.17, 76185 Karlsruhe, Germany} $^{ , 3}$ \and Abdolreza Nazemi \thanks{Institute of Information Systems and Marketing (IISM), Karlsruhe Institute of Technology, Kaiserstraße 12, 76131 Karlsruhe, Germany} \and Melanie Schienle \footnotemark[1]
	}

	\date{
		\today
	}
	\maketitle
	
	
	\vspace{-0.5cm}
	\thispagestyle{empty}
	\begin{abstract}
		\noindent
				
		We address challenges in variable selection with highly correlated data that are frequently present in finance, economics, but also in complex natural systems as e.g. weather. We develop a robustified version of the knockoff framework, which addresses challenges with high dependence among possibly many influencing factors and strong time correlation. In particular, the repeated subsampling strategy tackles the variability of the knockoffs and the dependency of factors. Simultaneously, we also control the proportion of false discoveries over a grid of all possible values, which mitigates variability of selected factors from ad-hoc choices of a specific false discovery level. In the application for corporate bond recovery rates, we identify new important groups of relevant factors on top of the known standard drivers. But we also show that out-of-sample, the resulting sparse model has similar predictive power to state-of-the-art machine learning models that use the entire set of predictors.
		
		\vspace{0,5cm}
		
		
		
		 \vspace{0.5in}
	\end{abstract}
	
	\footnotetext[3]{Corresponding author: Konstantin G\"{o}rgen; email: konstantin.goergen@kit.edu; phone: +49721-608-43793}
	
	\pagebreak

\section{Introduction}
In large-scale financial and economic but also complex natural systems as e.g. weather, with many potentially influencing variables for a target quantity, there is a key interest in detecting the relevant driving factors in a data-driven way. Such a fully data-adaptive choice of factors yields transparency in the identification of important channels avoiding biases from insufficient pre-specification but also inspiring and complementing future model building. 

We contribute to this literature by proposing a new knockoff-type methodology building on e.g. \citet{Candes2018} that offers control over the rate of falsely selected variables. With the false discovery rate (FDR) as the key hyperparameter, the selection and prediction performance based on the proposed technology is dominated by the FDR. It is thus transparent and interpretable while being data-driven since the FDR can be directly estimated as the empirical proportion of false discoveries (FDP). Note that this is in contrast to e.g. LASSO-type approaches \citep[see~e.g.][]{Tibshirani1996}, where penalty parameters can be chosen adaptively but have no stand-alone interpretation and meaning, which often creates a black-box connotation. Moreover, our technique gains robustness from simultaneously taking several nominal FDR-levels into account. In this way, we mitigate hyperparameter pre-selection effects and obtain robustness of results in the presence of time-dependent data. Both points are key for valid selection results in practice. We show that the proposed methodology provides interesting insights in detecting novel relevant factors for corporate bond recovery rates which might be important from a business but also regulatory perspective. In particular, we study the recovery rates of 2,079 U.S. corporate bonds that defaulted
between 2001 and 2016 depending on industry and stock specific information from Bloomberg Financial Markets and 144 macroeconomic market variables from the Federal Reserve Economic Data (FRED). For this, we also document superior out-of sample performance of the resulting sparse model using only relevant factors comparing them to state-of-the-art machine learning models on the entire and the selected set of predictors and to LASSO-type specifications. We confirm our point-wise ranking with results from model confidence sets \citep{Hansen2011}.

In particular, the proposed robustification technique works for the entire set of different knockoff baseline procedures from model-$X$ \citep{Candes2018} over deep knockoffs \citep{Romano2019} to group versions \citep{Dai2016} and mitigates the influence of hyperparameter input levels and data dependence challenges. We address the hyperparameter influence problem by proposing several weighted aggregation schemes for variable selection rates of different FDR-levels. By considering different weighting schemes, we account for vanishing scrutiny of the procedures in size of FDR-levels but extract the information from each level for the overall selection result. Secondly, we use a repeated subsampling scheme to control for the variability of the knockoff procedures, which themselves are random. While this shares similarities with \citet{Ren2021}, we employ subsampling \citep[see~e.g.][]{Meinshausen2010}, which provides robustness in the presence of correlated observations and high outliers. This is of key importance for the determination of relevant factors of the considered corporate recovery rates. In this empirical study, we additionally employ principal components analysis (PCA) on groups of macroeconomic variables of similar type to reduce cross-sectional correlation of the knockoff input factors while retaining interpretability on the group level. We investigate the performance of the proposed methodology for different baseline procedures and show that an ensemble yields superior out-of-sample prediction results. 

Our paper belongs to a recent quickly emerging literature that provides extensions and robustifications of the original model-$X$ knockoff framework \citep{Candes2018}, for example ultra-high-dimensional approaches employing pre-screening steps \citep{liu2022model,PAN2022107504} or resampling approaches focusing on reducing variability of the knockoffs \citep{Ren2021}. Other contributions diverge from the model-$X$ framework to provide better knockoffs for heavy-tailed distributions \citep{bates2021metropolized}, or by changing the process of generating the knockoffs, either via neural networks \citep{Romano2019} or by using different minimization functions \citep{spector2022powerful}.
Recently, a rapidly growing literature has also applied ML-based approaches in financial applications in particular for empirical asset pricing \citep{Chen_2019, Freyberger2020, Kelly2019}, where variable selection methods serve as a dimension reduction device and are mostly based on the lasso or other penalization frameworks with respective sparsity assumptions \citep{kozak2018,CHINCO_2019,FENG_2020,Freyberger2020}.
There has also been considerable research on recovery rates and loss given default (LGD), see e.g. \citet{Qi2011}, \citet{Leow2012}, 
\citet{Kaposty2020}, \citet{Kellner2022} for LGD and \citet{Nazemi2018,Nazemi2021}, 
and \citet{Bellotti2021} for recovery rates.

The rest of the paper is structured as follows. Section \ref{Sec:methods} introduces our proposed methodology in a general setting, while Section \ref{Sec:Application} focuses on the application. Therein, Section \ref{Sec:Data} introduces the data set of corporate bond recovery rates and Section \ref{Sec:Results} presents the main results of our analysis. Finally, we conclude in Section \ref{Sec:Conclusion}.

	
\section{Model Selection with Knockoffs} \label{Sec:methods} 
\subsection{Setup}
We propose a novel robustified knockoff-type procedure  \citep[e.g.][]{Candes2018} that offers direct control of the false discovery rate (FDR) while being robust to correlations in the data. The key hyperparameter FDR corresponds to the number of coefficients determined as non-zero while being truly zero relative to all obtained non-zero factors. FDR is well-known from the multiple testing literature as the type II error but has also been shown to directly link to the size of error rates in model estimation and prediction of after pre-selection with knockoffs \citep{Barber2019}. Thus setting the acceptable (nominal) FDR-level for knockoffs directly controls estimation and prediction performance of the resulting model, while the performance of other model selection techniques depends on parameters that lack a direct interpretation. Interpretable hyperparameters, however, are key for adequate tuning and eventual transparency of the results.  Moreover, the suggested knockoffs work irrespective of the type of underlying sparsity and in particular for high-dimensional cases, results are not dependent on a specific form of sparsity. We contribute a robust knockoff version which works in particular in the presence of strong cross-sectional and time dependence. This is key in many economic and financial applications and of peculiar importance for our application of the determination of relevant factors of corporate recovery rates.

We work in the following setting, where $y_i \in \mathbb{R}$ and  $X_i\in \mathbb{R}^{p}$ are observed for $i=1, \dots, n$ but only some unknown subset of the $p$ components in $X$ is relevant for $y$ and forms the so-called active set $\mathcal{S}$ of $X$, i.e. for $j\notin \mathcal{S}$, $y$ is independent of component $X_{(j)}$ conditional on $(X_{(k)})_{k\in \mathcal{S}}$. These components $\mathcal{S}$ should be selected in
\begin{align}
    y_i = f(X_i) + \epsilon_i \ ,
\end{align}
with an error term $\epsilon_i \in \mathbb{R}$ and a function $f(\cdot)$ that describes the impact of $X_i=(X_{i1},\dots,X_{ip})=(X_{ij},X_{i-j})$ for any $j=1, \dots p$ on $y_i$. In general, we assume that $f(X_i)=X_i \beta$, with $\beta\in \mathbb{R}^p$ for an easily interpretable structure, but we also include unknown non-parametric versions of $f$ in our application. We set as usual $Y=(y_1,\dots,y_n)'$ and $X=(X_1',\dots,X_n')'=(X_{(1)}, \dots,X_{(p)})= (X_{(j)},X_{(-j)})$ with $X_{(j)}\in\mathbb{R}^n$ for all $j= 1, \dots, p$.

In the literature, there exist different procedures for the construction of knockoffs such as model-$X$ knockoffs \citep{Candes2018}, deep knockoffs \citep{Romano2019}, and group-knockoffs \citep{Dai2016}. They all build on the same main idea to compare the regressors of interest $X$ with randomly generated knockoffs $\Tilde{X}=(\Tilde{X}_{(1)},\dots,\Tilde{X}_{(p)})$ that fulfill two properties:
\begin{enumerate}
    \item[(i)] pairwise exchangeability, i.e. the distribution of\\ $(X_{(j)}, X_{(-j)}, \Tilde{X}_{(j)}, \Tilde{X}_{(-j)})$ and $(\Tilde{X}_                    {(j)}, X_{(-j)}, {X}_{(j)}, \Tilde{X}_{(-j)})$ is identical
    \item[(ii)] $Y$ is independent of $\tilde{X}$ conditional on $X$.
\end{enumerate}
When regressing $Y$ on $(X,\Tilde{X})$ jointly, only those regressor components in $X$ which fundamentally differ from their corresponding ones in the random $\Tilde{X}$ according to a variable importance measure are judged as relevant and are part of the active set $\mathcal{S}$. The variable importance depends on model and estimation techniques, but for the linear case, e.g. the difference of absolute lasso coefficients of $X_{(j)}$ and $\Tilde{X}_{(j)}$ must be large enough.

For model-$X$ knockoffs \citep{Candes2018}, the two knockoff conditions (i) and (ii) are addressed by matching first and second moments of of $X$ and $\Tilde{X}$ in the construction of $\Tilde{X}$ subject to independence of $Y$ and $\Tilde{X}$ conditional on $X$. 
Matching expectations is straightforward and the second order construction leads to a convex optimization problem minimizing pairwise correlations of $X$ and $\Tilde{X}$ under the constraint that $Cov(X,\Tilde{X})$ be positive semi-definite. This effectively targets the off-diagonal elements of the covariance of $X$ and $\tilde{X}$ and leads to the approximate semi-definite program algorithm (ASDP) of \citet{Candes2018}. The obtained model-$X$ knockoffs $\Tilde{X}$ approximately fulfill conditions (i) and (ii), and the construction is exact if $(X,\Tilde{X})$ are normal. For a more detailed description of the construction of the more general deep knockoffs \citep{Romano2019} which fully operationalize the distributional form of (i) and (ii) and of group-knockoffs that use a pre-specified group-structure \citep{Dai2016}, see Appendix \ref{sec:Additional_Methods}. In general, the construction principle of all knockoff techniques is based on the standard Gaussian results in \citet{Barber2015} and the non-Gaussian, high-dimensional extension in \citet{Candes2018} to the model-$X$ knockoff filters. 

Once the knockoffs have been constructed, they can be used as a filtering device to select the active set. For this, each knockoff feature $\Tilde{X}_{(j)}$ is compared to its true counterpart ${X}_{(j)}$ via a feature-statistic $W_j$ for all $j=1,\dots, p$. In the linear case, for a lasso regression of $y$ on the joint $(X,\Tilde{X})$ over a grid of penalty parameters $\lambda$ with corresponding coefficients $\hat{\beta}_j(\lambda)$ we work with $\lambda_j=\sup{\{\lambda| \hat{\beta}_j(\lambda) \neq 0\}}$ as the largest $\lambda$ for which variable $j$ is in the active set and define 
\begin{align}
    W^{LCD}_j &= \vert \hat{\beta}_j (\lambda_0) \vert - \vert \hat{\beta}_{j+p} (\lambda_0) \vert \label{eq:lcd}\\
    W^{LSM}_j &= \sign(\lambda_j - \lambda_{j+p}) \max(\lambda_j,\lambda_{j+p})\label{eq:lsm} \ .
\end{align}
where $\lambda_0$ is chosen according to some global criterion like cross-validation and the $\sign(\cdot)$ function returns the sign of the input. Note that the $W_j$ from equations \eqref{eq:lcd} and \eqref{eq:lsm} correspond to the lasso coefficient difference (LCD) of the model-$X$ knockoffs and the lasso signed max (LSM) as described in \citet{Barber2015}, respectively. In practice, we mostly rely on the LCD measure which was shown to be preferable and robust to highly correlated features as in our application \citep{Candes2018}. Only for group knockoffs, we use the proposed adapted version of the LSM as suggested by \citet{Dai2016}. We only select variable components $j$ as part of the active set $\mathcal{S}$ if $W_j$ is greater or equal to some threshold $T$ with  
\begin{align}
     T=\min \left \{ t>0: \dfrac{\#\{j: W_j\leq -t\}}{\#\{j: W_j\geq t\}} \leq \alpha \right\} \ ,
\end{align}
where $\alpha \in [0,1]$ is the pre-specified level of acceptable (nominal) false discovery rate $FDR=\mathbb{E}[FDP]$, where $FDP=\dfrac{\vert \hat{S} \setminus {S} \vert }{\vert \hat{S}\vert}$ is the false discovery proportion, with $\hat{{S}}$ as the set of selected variables\footnote{In case $\hat{{S}}$ is empty, we set $FDR=0$ as in \citet{Candes2018}.} and $S$ as the set of truly relevant variables. In practice, we calculate this proportion as $\widehat{FDP}=\dfrac{\#\{j: W_j\leq -t\}}{\#\{j: W_j\geq t\}}$. Note that for the definition of $T$ we rely on \citet[Equation~3.9]{Candes2018} of the original suggestion of \citet{Candes2018}. 

\subsection{Robustified Knockoffs}
First, we propose an adapted version of the baseline knockoff procedures that can deal with time-dependence and an unknown, possibly non-standard covariate distribution due to high correlations among $X$. Secondly, we examine the full grid of possible nominal FDR-levels for the knockoff procedure to uncover dependence of selections on certain specific FDR-levels. We do this by repeating each robustified baseline procedure $K$ times over a grid of $K$ nominal FDR-values and combine the results by weighting the selection probabilities depending on the FDR-level. We call this procedure \textit{weighted FDR selection} (wFDR).\\
With covariate distributions different from normality and possible time-dependence, the standard assumptions from the Knockoff framework are violated, which could lead to strong variability of knockoff selections that are per definition random. We therefore suggest repeated subsampling to stabilize the selection procedure, motivated by the stability selection of \citet{Meinshausen2010} and \citet{Ren2021}, who suggest a similar procedure, where the knockoff procedure is repeated without subsampling.. We repeat the full knockoff procedure  $B=100$ times only using a subsample of the full $n$ observations, with subsampling rate $\theta$. The subsampling ensures that large outliers and data artifacts do not majorly affect the selection, while the repetition of the knockoff procedure controls the randomness of the knockoffs. This randomness would also allow no subsampling at all as in \citet{Ren2021}. For our application with a substantial amount of outliers in finite samples, however, we choose to use $\theta=0.9$. For a fixed FDR-level $\alpha$, the procedure ranks variables in decreasing order according to their selection frequency, i.e. empirical selection probability. The variable with the highest selection frequency receives rank $p$, the second most selected variable gets rank $p-1$, up to the least selected variable receiving rank 1. Alternatively, we also directly work with the selection probability instead of the ranks, which puts a larger emphasis on the variability of selection probabilities\footnote{Technically, it would also be possible to run a standard knockoff machine without subsampling and report either zero (no selection) or one (selection) for each variable. We refrain from such an approach due to the data challenges stated above.}. See also Method \ref{algo:stab_sel} for details. 
\begin{algorithm}
    \caption{Repeated Subsampling for Knockoffs}
  \begin{algorithmic}[1]
   \REQUIRE  
        \Statex Observation pairs $(X,Y)=(X_i,y_i)_{i=1}^n \in \mathbb{R}^{n\times{p+1}}$
        \Statex Nominal FDR-level $\alpha \in [0,1]$
        \Statex Knockoff procedure $Knock_{proc}$, e.g. model-$X$ knockoffs or deep knockoffs
        \Statex Subsampling rate $\theta$ and number of repetitions $B$
        
    \FOR{b in 1 to $B$}
        \STATE Draw random subsample $(X_b^{sub},Y_b^{sub})=(X_b^{s},y_b^{s})_{s=1}^{n_{sub}}$ (i.e. without replacement) of size $n_{sub}=\lfloor n\theta \rfloor$ 
        \STATE Apply $Knock_{proc}$ based on $(X_b^{sub},Y_b^{sub})$ and obtain $Ind_b=(ind_1^b,\dots,ind_p^b)$, where $ind_l^b$ is one if variable $l$ is selected and zero otherwise, $l=1,\dots,p$
    \ENDFOR
    \STATE Compute selection probability $pr_l$ for each variable $l=1,\dots,p$ as $pr_l=\dfrac{\sum_{j=1}^{B}ind_l^j}{B}$ 
    \ENSURE Selection probabilities for each variable  $P_{\alpha}=(pr_1,\dots,pr_p)$
  \end{algorithmic}
  \label{algo:stab_sel}
\end{algorithm}
Note that by construction the proposed subsampling adapted knock-off procedure keeps the fixed FDR-level $\alpha$ but robustifies the selection result.

Moreover, we propose to conduct each knockoff-baseline selection (Method~\ref{algo:stab_sel}) over a grid of $K$ different FDR values $\alpha_k$ jointly. Thus, for each fixed-level $\alpha_k$, we detect whether a variable is relevant or not and determine the corresponding selection probability via subsampling using our methodology.
Over the grid of different $\alpha_k$-values, the corresponding selection probabilities are then weighted depending on the level $\alpha_k$, where higher values of $\alpha_k$ receive lower weights corresponding to the definition of the FDR. The final selection probability for each variable is then obtained as the weighted sum of all selection probabilities for this component over all $\alpha_k$. Since the number of selected variables varies depending on the respective $\alpha$-level, our procedure prevents situations where results crucially depend on the pre-setting of one specific $\alpha$-level. We show explicitly that such situations happen in our empirical example where high correlations between variables exist and solve this issue by combining the results from distinct weighting schemes that control the influence of selections over the grid of possible FDRs. With that, we transparently control the FDR influence on selections while maintaining flexibility by not restricting the baseline procedure for selections too strongly. An overview of our method is given in Method \ref{algo:procedure}.
\begin{algorithm}
    \caption{Weighted FDR Selection}
  \begin{algorithmic}[1]
   \REQUIRE  
        \Statex Observation pairs $(X,Y)=(X_i,y_i)_{i=1}^n \in \mathbb{R}^{n\times{p+1}}$
        \Statex Set of nominal false discovery rates $FDR_k=\alpha_k \in [0,1]$, $k=1,\dots,K$
        \Statex Baseline procedure $B((X,Y),FDR_k)$ that returns selection probabilities for $X$ given $FDR_k$, e.g. as in Method \ref{algo:stab_sel}
        \Statex Weighting scheme $\omega=(\omega_1,\dots,\omega_K)$ that assigns a weight depending on selection run $k$
    \FOR{k in 1 to K}
        \STATE Run baseline procedure $B(X,Y,FDR_k)$ for $FDR_k$ and obtain selection probabilities $P_k=(pr_{1k},\dots,pr_{lk}$) for each variable $l=1,\dots,p$ \LineComment{Possible $P_k$ format: 0/1-coding, rank, probabilities, see Method \ref{algo:stab_sel} for computation}
        
    \ENDFOR
     \FOR{l in 1 to p}
        \STATE Obtain weighted selection probability $WP_l=\sum_{k=1}^K pr_{lk} \omega_k $
    \ENDFOR
    \ENSURE Final weighted selection probabilities for each variable  $WPr=(WP_1,\dots,WP_p)$ 
  \end{algorithmic}
  \label{algo:procedure}
\end{algorithm}
We suggest two different weighting schemes that all depend on the following observation. By definition, a low nominal FDR implies that the number of falsely selected variables is small compared to the number of selected variables. This suggests that weighting should be conducted in a way that low FDRs, for which selection probabilities are thus more informative, should receive higher weight. Imagine we have two selection probabilities $pr_j^{low}$ and $pr_k^{high}$ for variable $j$ and $k$ at nominal levels $\alpha_{low}=0.1$ and $\alpha_{high}=0.95$, respectively. For $\alpha_{low}$, less than $10\%$ selected variables should be false selections, while for $\alpha_{high}$ less than $95\%$ of selections should be false. Obviously, when $pr_j^{low}$ and $pr_k^{high}$ are very similar, one would give variable $j$ a higher weight in being a true influencing variable compared to variable $k$.  To formalize this intuition, we propose two weighted averages and compare them with an unweighted baseline. One average is just based on weights that decay linearly, while the other one uses weights that decay exponentially and are equidistant on the log-scale. More specifically, for the value (probability or rank) at $FDR_k=\alpha_k$, where $k=1,\dots, K$, and $FDR_k \in (0,1)$ on an equidistant grid, the linear weight for position $k$ on the FDR-grid is given by
\begin{align}
  \omega_k^{lin}=\dfrac{K-k+1}{\sum_{k=1}^K k}  \ ,
\end{align}
and the exponentially decaying weight is given by
\begin{align}
    \omega_k^{exp}=\dfrac{\exp{\left(ln(K)-(k-1)\dfrac{ln(K)-ln(1)}{K-1}\right)}}{\sum_{k=1}^K \exp{\left(ln(K)-(k-1)\dfrac{ln(K)-ln(1)}{K-1}\right)}} \ .
\end{align}
We compare these weights with an unweighted average, where we expect the weighted averages to be more informative and thus give better indications of true influencing variables than their unweighted counterparts.


\section{Empirical Study: Corporate Recovery Rates} \label{Sec:Application}


\subsection{Data} \label{Sec:Data}

Our empirical study uses a data set consisting of 2,079 U.S. corporate bonds that defaulted between 2001 and 2016 obtained from S\&P Capital IQ-similar. We retrieved industry and stock variables from Bloomberg Financial Markets. Moreover, we collected 144 macroeconomic variables that were used in previous credit risk studies from the Federal Reserve Bank of St. Louis (FRED, Federal Reserve Economic Data). We classified these macroeconomic variables into 20 groups as detailed in Appendix C. 
We structured the groups according to financial conditions (Loans, Bank Credit and Debt), monetary measures (Savings, CPIs, Money Supply), corporate measures (Cash Flow and Profit), business cycle (Unemployment, Industrial Production, Private Employment, Housing, Income, Real GDP, Inventories), stock market (Index Returns and Volatilities), international competitiveness (Exchange Rates, Trade), and micro-level factors (Producer Price Index). These groups are tailored to yield interpretable factors and are more granular than in \citet{Nazemi2021}, who consider a prediction-focused analysis. As can be seen from Table \ref{tab:correlations_groups}, variables within groups are often highly correlated, which makes it hard to directly analyze them without transforming the data. Although there are still a few highly dependent groups, the correlation between groups is much smaller in general, which can be seen in Figure \ref{tab:correlations_groups}. There, the median correlation across groups is shown, which the diagonal indicating the median correlation in-group, corresponding to column $50\%$ in the Table on the left of Figure~\ref{tab:correlations_groups}. \\

\begin{table}
\caption{Summary Statistics of the Distribution of Pairwise Correlations: Detailed Within and Schematic Cross-Group}
\begin{minipage}{10cm}
\centering
\resizebox{\textwidth}{!}{%
\begin{footnotesize}
\begin{tabular}{|lrrrrr|r|}
  \hline
Group & Min & 25\% & 50\% & 75\% & Max & \# members\\ 
  \hline
  	1: Financial Conditions: Loans & 0.50 & 0.78 & 0.92 & 0.97 & $1.00^*$ & 6 \\ 
	2: Monetary Measures: Savings & 0.14 & 0.29 & 0.43 & 0.67 & 0.91 & 3 \\ 
	3: Monetary Measures: CPIs & -0.91 & -0.17 & 0.63 & 0.94 & $1.00^*$ & 13\\
	4: Monetary Measures: Money Supply & 0.90 & 0.91 & 0.93 & 0.96 & $1.00^*$ & 4 \\ 
	5: Corporate Measures: Cash Flow and Profit & 0.40 & 0.66 & 0.76 & 0.88 & 0.94 & 4\\
	6: Business Cycle: Unemployment & -0.29 & 0.51 & 0.78 & 0.92 & $1.00^*$ & 10 \\ 
	7: Business Cycle: Industrial Production & -0.54 & 0.20 & 0.51 & 0.83 & 0.98 & 13 \\ 
	8: Business Cycle: Private Employment & -0.26 & 0.20 & 0.59 & 0.79 & 0.98 & 10 \\ 
	9: Business Cycle: Housing & 0.68 & 0.91 & 0.96 & 0.97 & 0.99 & 12\\ 
	10: Business Cycle: Income & -0.32 & -0.28 & -0.21 & 0.48 & 0.99 & 4\\ 
	11: Stock Market: Index Returns and Volatilities & -0.66 & -0.40 & -0.20 & 0.40 & 0.99 & 9\\ 
	12: International Competitiveness: Exchange Rates & -0.66 & -0.26 & 0.68 & 0.81 & 0.97 & 5\\ 
	13: International Competitiveness: Trade & -0.81 & -0.65 & -0.43 & 0.34 & 0.96 & 5\\ 
	14: Micro-level: Bond Yields and Rates & -0.90 & -0.40 & 0.34 & 0.86 & 1.00 & 20 \\ 
	15: Micro-level: Bond Defaults in Industry & - & - & - & - & - & 1 \\
	16: Micro-level: High Yield Default Rate  & - & - & - & - & - & 1\\
	17: Financial Conditions: Bank Credit and Debt & -0.87 & -0.17 & 0.45 & 0.90 & $1.00^*$ & 11 \\ 
	18: Business Cycle: Real GDP & 0.35 & 0.47 & 0.59 & 0.74 & 0.89 & 3 \\ 
	19: Micro-level: Producer Price Index & 0.81 & 0.92 & 0.96 & 0.98 & $1.00^*$ & 6 \\ 
	20: Business Cycle: Inventories & 0.29 & 0.32 & 0.59 & 0.83 & 0.98 & 4 \\ 
   \hline
   \multicolumn{7}{l}{\footnotesize *These values are only due to rounding.}
\end{tabular}
\end{footnotesize}
}
\end{minipage}
\begin{minipage}{6.5cm}
    \centering
    \includegraphics[width=\textwidth]{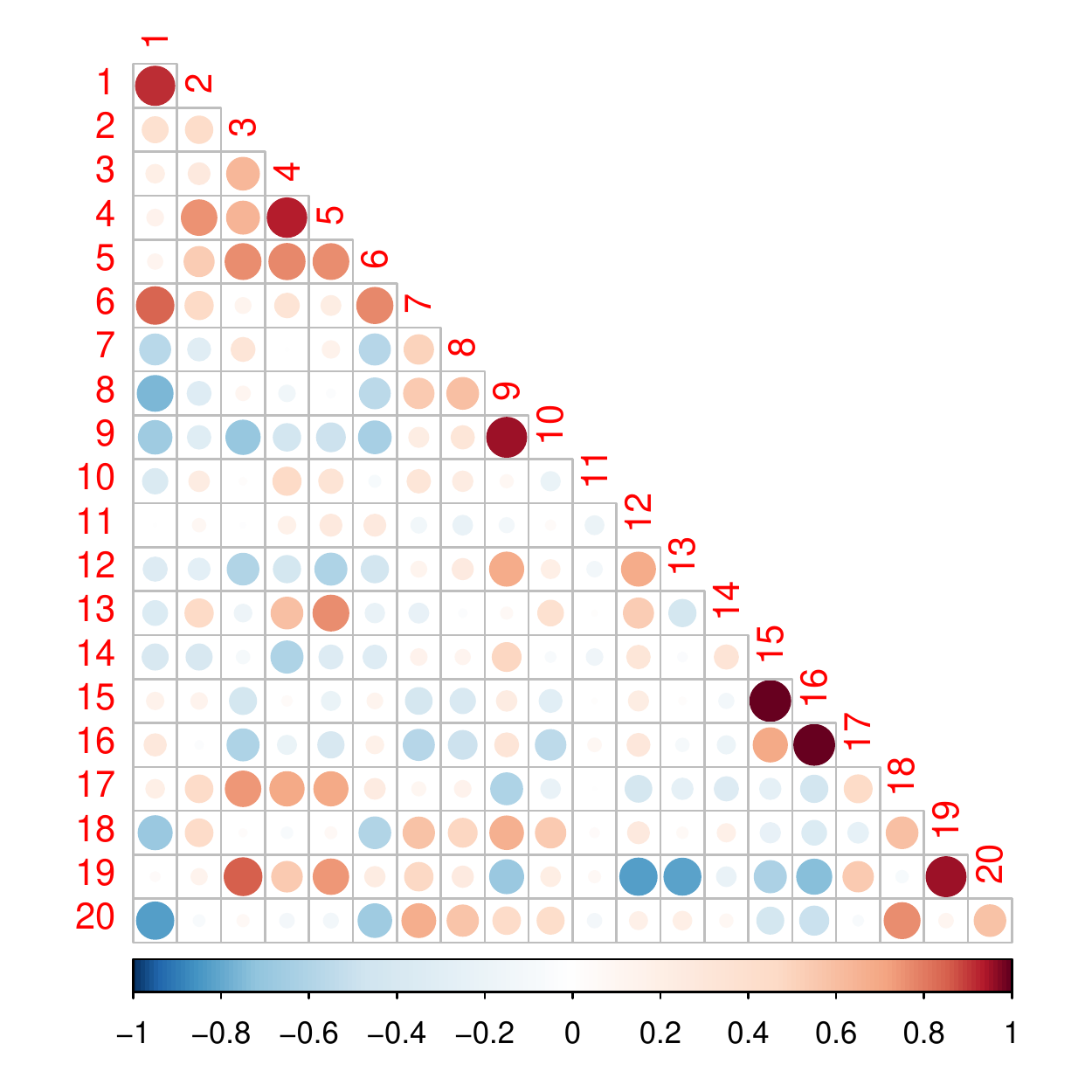}
\end{minipage}
\caption*{In the table on the left values depict summary statistics of the distribution of pairwise correlations within each group. The median can be found on the diagonal of the schematic figure on the right which moreover displays the median of all cross-group correlations. Details on the variable components of each group are listed in Table~\ref{tab:groups}.}\label{tab:correlations_groups}
\end{table}

\begin{figure}
		\centering
		\includegraphics[width=0.7\textwidth]{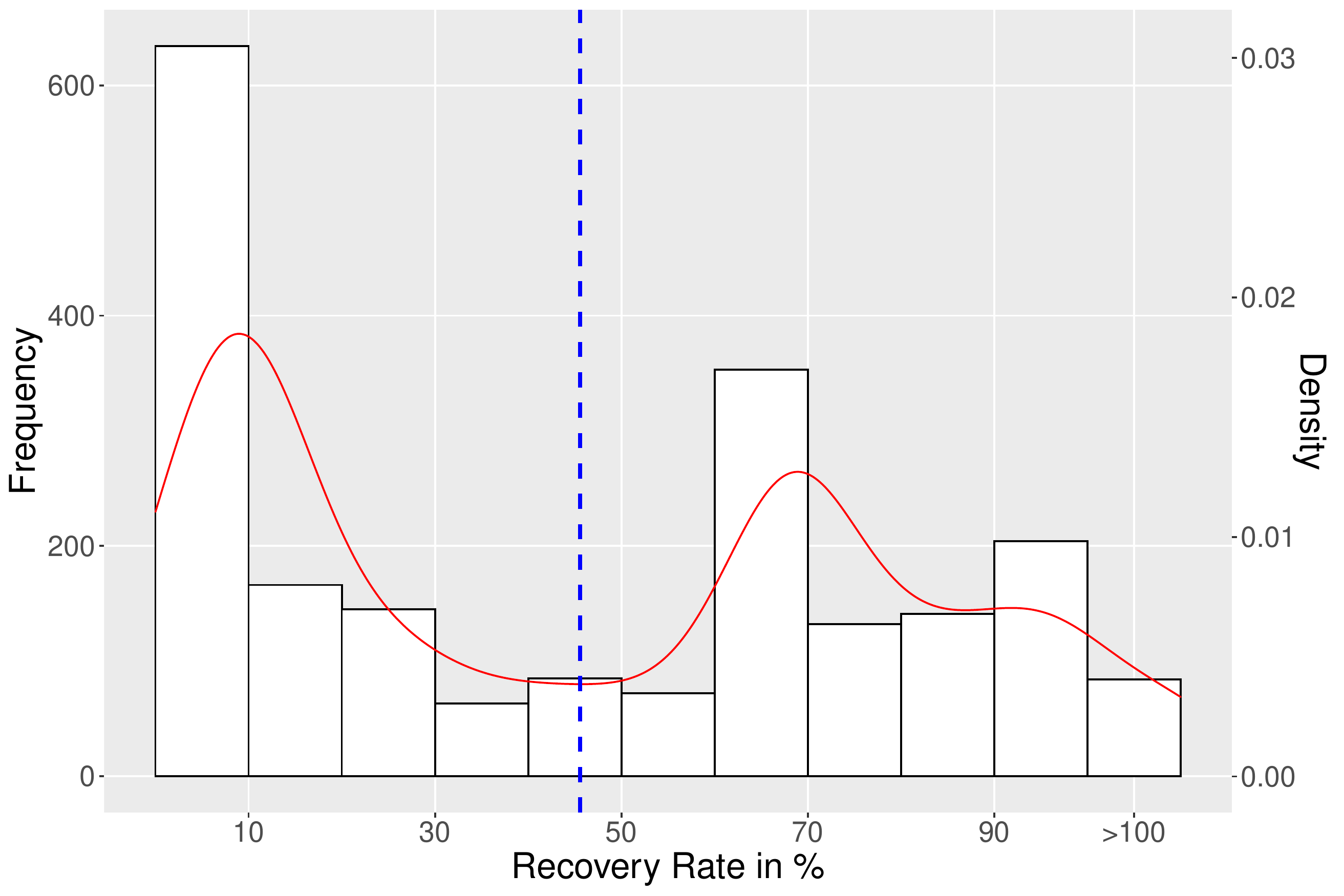}
		\caption{Recovery rate frequency and density (red) for the defaulted US corporate bonds from 2001 to 2016 and mean recovery rate in dashed line (blue).}
		\label{fig:RR_density}
	\end{figure}

The recovery rate is defined as the mean of trading price between the default day and data 30 days after default, which we retrieve from Capital IQ. The data is originally from the Trade Reporting and Compliance Engine (TRACE). In our analysis, all corporate bonds have debt values, at the time of default, of greater than \$50 million. The mean value of the recovery rate for the 2,079 U.S. corporate bonds in our sample is 45.57 percent, and the sample standard deviation is 35.04 percent. The empirical distribution of the recovery rates of defaulted US corporate bonds naturally peaked in the financial crisis from 2008-2010\footnote{A more detailed figure on the distribution of defaults over time can be found in Figure \ref{fig:Defaults_hist_plot} in Appendix \ref{Sec:Fig_Tables}. A large share of defaults was caused by both the Lehman Brother bankruptcy in September 2008, and the CIT Group Inc. bankruptcy in November 2009.}. Around 30 percent of defaulted bonds have recovery rate less than 10 percent. There is another distribution peak in the range of values between 60 percent and 70 percent, which is visualized in Figure \ref{fig:RR_density}.

In addition, our bonds consist of four seniority levels for the bonds: (i) senior secured, (ii) senior unsecured, (iii) senior subordinated, (iv) subordinated, and junior subordinated. An overview over the distributions over the different bond types can be found in Appendix \ref{Sec:Fig_Tables} in Figure \ref{fig:RR_bond_classes}. Since most defaults (82.5\%) occurred in the class of senior unsecured bonds, which is driving the distribution of recovery rates, we decided to not distinguish between groups of seniority levels in our analysis. Additionally, the sample size would be too small for such a sub-analysis.

\subsection{Empirical Results} \label{Sec:Results}
In this subsection, we use our methodology to identify and quantify the recovery rates of corporate bonds in a data-driven way. This is key in practice for investments, hedging, and supervision but also for model building and interpretation. As an extension in a comprehensive out-of-sample forecasting study, we also demonstrate that simple linear predictions based on variables selected by the knockoff procedures can compete with and often improve upon non-parametric methods that employ the full set of variables. Moreover, we show that the obtained knock-off selection of variables is robust to discarding certain time subperiods, e.g. after the financial crisis.

\subsubsection{Identification of Important Groups and Effects on Recovery Rates} \label{sec:results_identification}
To determine the driving factors of corporate bond recovery rates, we use our proposed methodology on the full sample from 2001 to 2016 for the data-driven selection of relevant components. We show results of our suggested combined subsampling and weighted FDR selection technique (see Method \ref{algo:stab_sel} and \ref{algo:procedure}) across all types of different baseline-knockoff methods, i.e. in particular, we study model-$X$ knockoffs, deep knockoffs with two different neural network architectures, and group knockoffs. 

Since our data is highly correlated within groups (see Table \ref{tab:correlations_groups}) and we are mainly interested in group effects and selections, we transform our data using principal component analysis (PCA)\footnote{See e.g. \citet[Chap.~14.5]{Hastie2009}.}. To retain interpretability on a group level, we conduct one PCA per group and only use the most important principal components (PC) to describe that specific group, i.e. a maximum of four PCs that explain at least 90\% of the variability in the group. This helps to reduce high correlations among variables and break down large groups of variables to one or two components to see their main effects, avoiding multicollinearity issues in post-selection linear models. Additionally as a robustness check for the model-$X$ knockoffs, we employ a smaller version using a maximum of two PCs ("2comp").This group-PCA step greatly reduces the dimensionality of the data and serves as a viable alternative to other pre-screening procedures such as omitting variables with high pairwise correlations. With that, the variables are scaled by their standard deviation and centered around zero. Similar approaches in reducing dimensions have also been taken by \citet{Kelly2019} in an asset pricing application.

\begin{figure}
	\centering
	\includegraphics[width=0.9\textwidth]{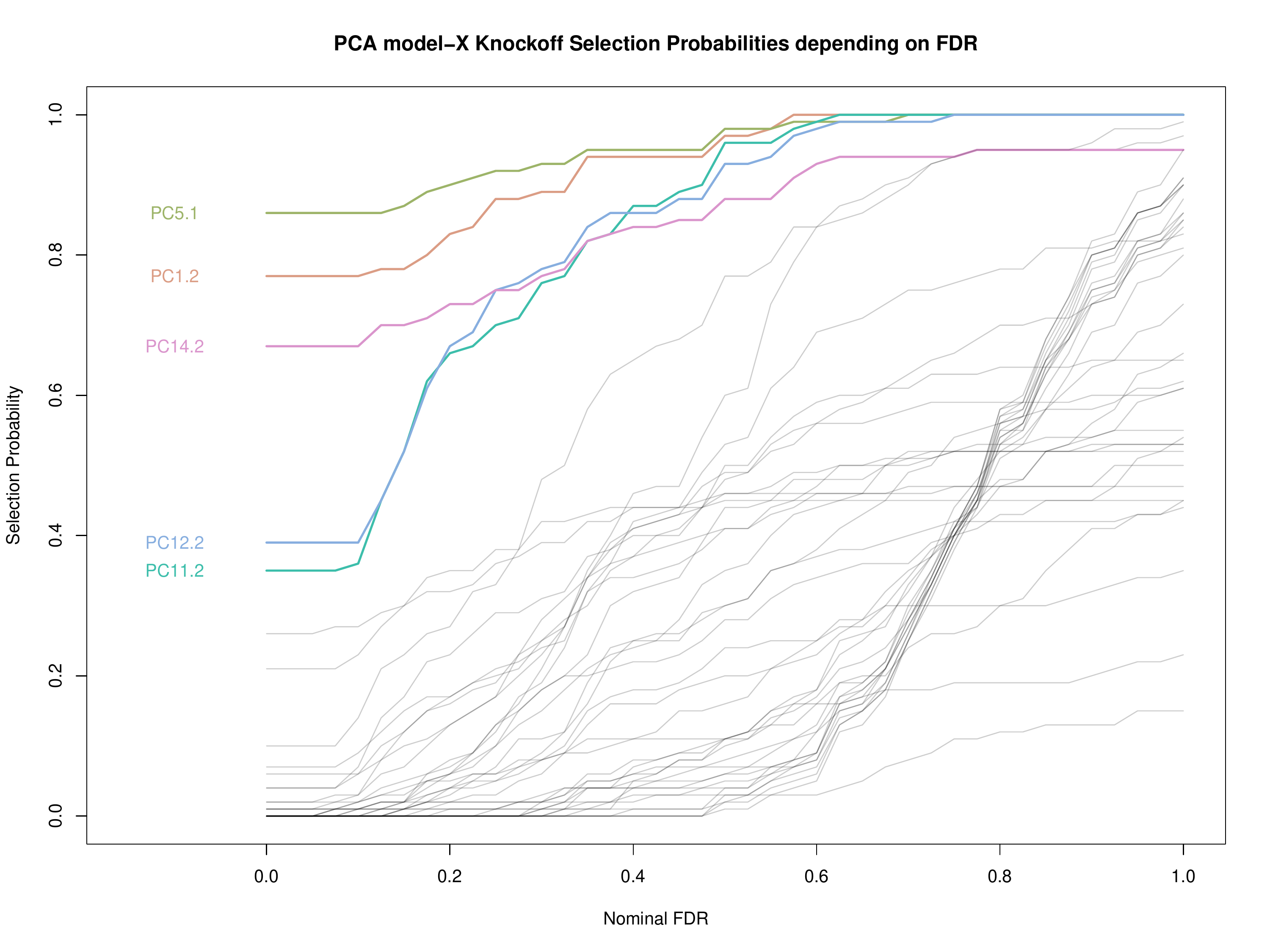}
	\caption{model-$X$ Knockoff Selection Probabilities for different nominal FDR using group principal components. Selection Probabilities are obtained rerunning the full knockoff procedures using repeated subsampling of 90\% of the data (100 iterations). Highlighted groups have the highest mean selection rank, i.e. the mean over the rank in each FDR-scenario. The PCA component with the highest probability receives the highest rank ($=41$) and vice versa ($=1$).}
	\label{fig:PCA_Selection_probs_FDR}
\end{figure}

Figure \ref{fig:PCA_Selection_probs_FDR} graphically shows the most important PCA-features for model-$X$ knockoffs over all possible nominal FDRs, while similar figures for the other procedures can be found in Appendix \ref{Sec:Fig_Tables} (Figure \ref{fig:selection_probs_add_FDR}). Most prominently, the selection probabilities of important features are rather high (with selection frequency of 0.6) already at a nominal $FDR=0.2$ and rise to 0.8 at nominal $FDR=0.4$ for the model-$X$ procedures, where other, less important factors only attain similar levels from a nominal $FDR=0.8$ onwards. Such levels of FDR are clearly undesirable, but investigating the entire grid of FDR-levels jointly and with appropriate weighting is beneficial and yields robustness due to the rather high variability of selection probabilities for minimal changes at a considered specific nominal FDR-level. For the deep knockoff procedures, however, we see high selection probabilities for relevant features throughout all FDR levels (see Figure \ref{fig:selection_probs_add_FDR} (bottom)). Comparing different structures for the neural networks in the deep knockoffs, this effect is more pronounced for narrower networks with only 5 neurons per layer. Since a wider network can learn more complex structures, it can build more accurate knockoffs and with that, identify variables that are less likely to be true influencing variables, especially for cases where nominal FDR-levels are low. This highlights the importance of considering multiple methods for generating knockoffs and combining their insights to identify the most important groups.

\begin{figure}
	\centering
	\includegraphics[width=0.95\textwidth]{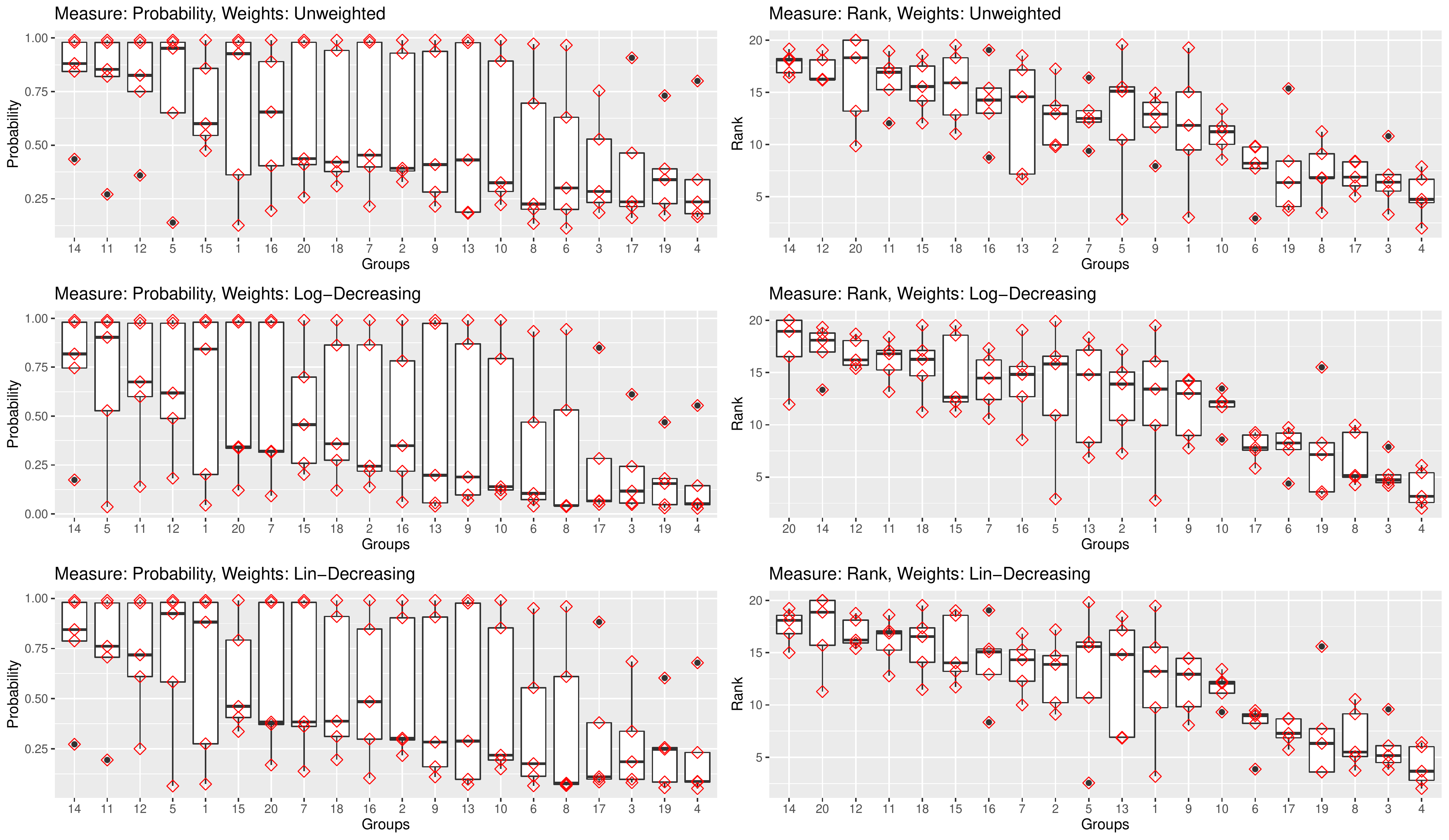}
	\caption{Boxplots of weighted mean selection probabilities/ranks of each group using different procedures. In each subplot, the red squares represent the means of each procedure. Groups are sorted by highest probability/rank from left to right. For PCA procedures, the group probability/rank is assigned as the highest value over all group-principal components. In case of ranks, the PCA-ranks are rescaled linearly to lie between 1 and 20.}
	\label{fig:selection_probs_rankings}
\end{figure}

To finally select appropriate variables over all five methods and FDRs, we compare two different ranking procedures for variables at each FDR-level. To combine the different rankings from each FDR-level to a final selection (probability), we employ three distinct weighting approaches. First, we distinguish between using ranking of variable selections (from 20 to one) or selection probabilities from our procedures. The subsequent weighting of ranks/probabilities for each variable over all nominal FDR-values is conducted as in Section \ref{Sec:methods} using either equal weighs, linear decaying weights ($\omega_k^{lin}$, \textit{Lin-decreasing}), or exponentially decaying weights ($\omega_k^{exp}$, \textit{Log-decreasing}). The weights assign the highest weights to low FDR-values and decrease with higher FDR-levels (see Section \ref{Sec:methods} for details). Figure \ref{fig:selection_probs_rankings} shows boxplots over selections from different methods by both ranking procedures and the three weighting schemes. In general, we can see that group 14, 11, and 12 are always among the top four in each procedure, while group 5 only appears as important when looking at probabilities, and group 20 vice versa only when looking at ranks. Table \ref{tab:selections_method_prob_rank} highlights the influence selection probabilities and ranks, where both group 5 and group 20 are given higher relative importance in the weighted schemes that focus on low nominal FDRs. Otherwise, results for the most selected groups are mostly stable over both schemes and all weights. This is largely in line with the literature that also determines the factors in group 14, 11, 20, and 5 as relevant with a more simplistic and less robust data-driven selection technology \citep{Jankowitsch2014,Nazemi2018}\footnote{Other, less important groups that have been selected often by our approach include high yield default rates, defaults in the respective industry, and GDP measurements. These findings are also in line with \citet{Nazemi2018} and \citet{Jankowitsch2014}, who report similar factors to be important.}. Though different from the existing studies, however, our methods additionally also detect group 12 as important, that consists of exchange rates. 

\begin{table}[ht]
\centering
\caption{Most-Selected Groups Over Different Weighting Schemes} 
\label{tab:selections_method_prob_rank}
\resizebox{0.9\textwidth}{!}{%
\begin{tabular}{lrrrrrrrr}
  \hline
  Rank: & \multicolumn{2}{c}{1} & \multicolumn{2}{c}{2} & \multicolumn{2}{c}{3} & \multicolumn{2}{c}{4} \\
  \hline
 & Group & Mean Score & Group & Mean Score & Group & Mean Score & Group & Mean Score \\ 
  \hline
Prob\_unweight & 14 & 0.826 & 11 & 0.783 & 12 & 0.781 & 5 & 0.742 \\ 
  Prob\_exp & 14 & 0.741 & 5 & 0.687 & 11 & 0.676 & 12 & 0.651 \\ 
  Prob\_lin & 14 & 0.775 & 11 & 0.726 & 12 & 0.709 & 5 & 0.708 \\ 
  Rank\_unweight & 14 & 17.768 & 12 & 17.156 & 20 & 16.276 & 11 & 16.100 \\ 
  Rank\_exp & 20 & 17.479 & 14 & 17.300 & 12 & 16.808 & 11 & 16.139 \\ 
  Rank\_lin & 14 & 17.547 & 20 & 17.172 & 12 & 16.880 & 11 & 16.117 \\ 
   \hline
\end{tabular}
}
 \caption*{\footnotesize Note: In the columns, \textit{Group} depicts the selected variable group, while \textit{Mean Score} shows the (weighted) mean over all five procedures for the four most selected groups. $Prob$ and $Rank$ refer to whether probabilities or ranks are used, while $unweight$, $exp$, and $lin$ refer to the weighting scheme of equal weighting, linear-decreasing weighting, and exponentially decreasing weighting. The ranks for groups are rescaled linearly to lie between 1 and 20 (20 being the best score), while the selection probabilities lie between 0 and 1.} 
\end{table}

Note that group 14 describes bond yields of major bonds and rates of different general indicators such as mortgage, treasury, and loans, which might be considered naturally predictive for the state of the economy and thus of bonds. The data-driven selection therefore confirms the intuition that these indicators have an influence on recovery rates. Similarly, the selection of the other groups can be explained. These describe international exchange rates against the USD (group 12) and stock market indicators such as returns and volatilities of the most important indices (group 11). Furthermore, capacity utilization of industries and change in inventories of private and businesses play an important role (group 20), as well as corporate measures such as profit of the firm and cash flow (group 5).
More specifically, in the light of the financial crisis in 2008 and the following euro crisis, exchange rates were strongly affected \citep[see~e.g.][]{McCauley2009,Kohler2010}, which might explain their connection to recovery rates in regard to globally active firms. The capacity utilization group, on the other hand, measures how much of total potential output is actually utilized by industry and additionally contains information about inventories (and their change over time). This is highly relevant for recovery rates when thinking of a firm's business model in general and inventories of firms that could indicate how much can be recovered given default. Naturally, returns and volatilities of the most important indices such as the S\&P 500 or the NASDAQ 100 describe the general situation of the economy and the value of companies, which again is an indicator of recovery rates of these firms. Finally, profit and cash flow are probably the most direct factors for short-term companies finances, and are thus a good predictor for the default of a firm. 

As a robustness check, we also computed variable importance measures from a random forest model using mean variance reduction as a measure for the importance of a group\footnote{To give the nonparametric random forest maximum flexibility, we used the the raw data as input instead of using the aggregated PCA-groups.}. Computational details of this model can be found in Appendix \ref{sec:Random_forest}. Table \ref{tab:rf_altmann_importance} in Appendix \ref{Sec:Fig_Tables} shows the mean values aggregated on a group level of variance reduction and corresponding p-values from the PIMP procedure of \citet{Altmann2010}. There, we can confirm that especially group 12, 14, and also 11 have high importance, while group 20 and group 5 are not deemed important. This can be explained by the predictive nature of the measure and method that favors groups such as group 15, which measures the bond defaults within the industries. Interestingly, group 15 is also in the top 6 groups of many of the other procedures, although mostly ranked below all the other selected groups.

\subsubsection{Post-Selection Performance}

To obtain an unbiased quantification of the effect of each factor, we re-estimate a linear model using only those groups that were selected in Section \ref{sec:results_identification}. We show the effect of the most important principal components (PCs) for each group in Table \ref{tab:pca_lm_groups}. While the effects of selected variables appear mostly significant, it has to be noted that assuming a linear model might be too optimistic and effects between groups might be affected by some remaining multicollinearity between similar groups.

Using the PCs allows using and working with the strong correlation within groups, and facilitates the interpretation and comparison of effects between different groups since variables in each group are centered and scaled. The results of Table \ref{tab:pca_lm_groups} show that most of the selected coefficients seem highly significant, where focus should lie on the first and second PC which capture the largest share of the variance in each group. Group 14 has a largely negative impact on recovery rates, although especially the last PC is affected by inclusion of more variables switching signs. Group 11 has a primarily positive impact (in the first two PCs), while adding large explanatory power (Adjusted $R^2$), It is, however, also correlated with group 12, which adds little explanatory power, but also has a significant positive first PC. Group 20 and 5 have a negative impact but appear to affect the coefficients of other components, which is why we consider their inclusion rather as a robustness check, since they also do not add as much to an increase in $R^2$ as for example group 14 and 11. Group 12 is a special case, having both positive (PC1) and negative (PC2) significant coefficients. Taking a closer look at the weights of the PCs\footnote{A complete list of the PCA-weights can be found in Table \ref{tab:pca_weights} in the Appendix.}, the first PC assigns a large negative weight to the exchange rates of Canadian dollar, Swiss franc against one USD, and the real broad effective exchange rate for the US, while the second PC gives large negative weight to the rate of USD against one British pound. 

\begin{table}[!htbp]
  \centering 
	\caption{Linear Regression with PCA-Components of Most-Selected Groups} 
	\label{tab:pca_lm_groups}
	\resizebox{0.9\textwidth}{!}{%
\begin{tabular}{@{\extracolsep{5pt}}lccc|cc} 
\multicolumn{4}{c}{Most Selected Groups} & \multicolumn{2}{c}{Additional Groups}
    \\\hline & & & & & \\[-1.8ex] 
    & Group 14 & Group 14,11 & Group 14,11,12 & Group 14,11,12,20 & Group 14,11,12,20,5\\ 
	\hline & & & & &\\[-1.8ex] 
  PC5.1 &  &  &  &  & $-$14.225$^{***}$ (2.456) \\ 
  PC5.2 &  &  &  &  & 1.751 (4.604) \\ 
  PC11.1 &  & 1.895$^{***}$ (0.544) & $-$0.973 (0.646) & $-$0.922 (0.645) & 0.467 (0.651) \\ 
  PC11.2 &  & 9.386$^{***}$ (0.594) & 8.524$^{***}$ (0.675) & 8.782$^{***}$ (0.784) & 8.841$^{***}$ (0.717) \\ 
  PC11.3 &  & $-$4.464$^{***}$ (1.150) & $-$1.316 (1.242) & 0.285 (1.474) & $-$1.337 (1.530) \\ 
  PC11.4 &  & $-$8.865$^{***}$ (1.426) & $-$1.936 (1.737) & 0.232 (2.205) & $-$1.551 (2.274) \\ 
  PC12.1 &  &  & 3.953$^{***}$ (0.706) & 2.925$^{***}$ (0.738) & 4.402$^{***}$ (0.748) \\ 
  PC12.2 &  &  & $-$6.158$^{***}$ (1.387) & $-$7.384$^{***}$ (1.663) & $-$7.530$^{***}$ (2.035) \\ 
  PC14.1 & $-$1.346$^{***}$ (0.205) & $-$3.428$^{***}$ (0.309) & $-$1.826$^{***}$ (0.499) & $-$2.032$^{***}$ (0.562) & $-$0.694 (0.595) \\ 
  PC14.2 & 3.254$^{***}$ (0.363) & 3.312$^{***}$ (0.605) & 3.888$^{***}$ (0.636) & 4.201$^{***}$ (0.995) & 8.411$^{***}$ (1.062) \\ 
  PC14.3 & $-$1.352$^{**}$ (0.567) & 0.238 (0.992) & $-$3.044$^{***}$ (1.016) & $-$2.684$^{*}$ (1.467) & $-$2.816$^{*}$ (1.501) \\ 
  PC20.1 &  &  &  & 1.373 (1.181) & $-$0.643 (1.991) \\ 
  PC20.2 &  &  &  & $-$5.211$^{***}$ (1.769) & $-$18.210$^{***}$ (4.366) \\ 
  Constant & 45.574$^{***}$ (0.748) & 45.574$^{***}$ (0.691) & 45.574$^{***}$ (0.682) & 45.574$^{***}$ (0.681) & 45.574$^{***}$ (0.672) \\
 \hline & & & & & \\[-1.8ex]  
Observations & 2,079 & 2,079 & 2,079 & 2,079 & 2,079 \\ 
R$^{2}$ & 0.056 & 0.170 & 0.188 & 0.223 & 0.244 \\ 
Adjusted R$^{2}$ & 0.055 & 0.168 & 0.185 & 0.219 & 0.239 \\ 
Residual Std. Error & 34.070 (df = 2075) & 31.958 (df = 2073) & 31.637 (df = 2071) & 30.964 (df = 2067) & 30.574 (df = 2065) \\ 
F Statistic & 41.139$^{***}$ (df = 3; 2075) & 85.117$^{***}$ (df = 5; 2073) & 68.351$^{***}$ (df = 7; 2071) & 54.053$^{***}$ (df = 11; 2067) & 51.147$^{***}$ (df = 13; 2065) \\ 
\hline \\[-1.8ex] 
\end{tabular}
}
 \caption*{\footnotesize Note: PCX.Y stands for principal component Y of group X. Variables were selected taking all groups among the four most-selected groups over all weighting schemes. See Table \ref{tab:selections_method_prob_rank} for details on group selection. We include PCs in each group until they explain more than 90\% of total variability in that group. Coefficients are shown with stars according to their significance in t-tests. SEs in parentheses are HC3 robust. $^{*}$p$<$0.1; $^{**}$p$<$0.05; $^{***}$p$<$0.01.} 
\end{table}
It is in line with intuition that higher bond yields and rates such as mortgages have a negative impact on recovery rates (group 14), since they might indicate a riskier environment. The positive impact of group 11 can be attributed to the fact that higher returns and smaller volatilities in the stock indices result in larger recovery rates. At the same time, the positive impact of exchange rates (group 12) in the first PC indicates that when the USD is weak against other major currencies, recovery rates are higher, while the opposite effect is observed in PC2. This effect could be explained by defaulted companies holding assets in foreign currencies that are more valuable when the USD is weak. On the other hand, we cannot fully rule out that this effect is caused by the USD exchange rate dropping against other major currencies \citep[see~e.g.][]{Kohler2010} because of large crash events that are connected to bond defaulting. 

\subsubsection{Extension: Out-of-Sample Prediction Performance}
In addition to the identification and interpretation of important factors explaining recovery rates, we also assess the out-of-sample forecasting performance of the reduced models in various scenarios. 
Here, we distinguish between two main cases: firstly, we check the infeasible forecasting scenario as reference point, where we use the determined models from Section \ref{sec:results_identification} employing information from the full data comprising 2001-2016 in the model selection step when forecasting for the year 2012-2016 (Table \ref{tab:oos_results_full}). Additionally, we also provide results for the ``completely'' out-of-sample forecasting case , where we re-determine all models on a limited time period from 2001-2011 and predict 2012-2016 (see Table \ref{tab:oos_results_short}). For the post-selection estimation step, we use a wide variety of models ranging from simple linear methods, standard and penalized, up to flexible fully non-parametric methods such as random forests (see Appendix \ref{sec:Random_forest} for implementation details). Moreover, we consider both cases with the full raw data and group-PCA-transformed data in different settings. 
In all settings, we clearly confirm that knockoff pre-selection improves prediction performance. The results also highlight that this cannot be achieved with lasso pre-selection, thus confirming the importance of our robust approach. After knockoff pre-selection, simple (penalized) linear forecasting models often achieve quite competitive forecasting performance with only slight improvements by a non-linear fit. This highlights that forecasting with a data-driven selection of important predictors pays-off, while maintaining easy interpretation in comparison to their fully non-parametric counterparts.

We assess the forecasting performance by calculating the root-mean-squared forecasting error $RMSE=\sqrt{\dfrac{1}{K}\sum_{\tau=k}^T (\hat{y}_\tau-y_\tau)^2}$ and the mean-absolute error $MAE=\dfrac{1}{K}\sum_{\tau=k}^T \vert\hat{y}_t-y_\tau \vert$ for a prediction $\hat{y}_\tau$ of $y_\tau$ at forecasting time $\tau=k,\dots,T$, and forecast length $K=T-k$. We use different forecast constructions with fixed, expanding and rolling windows on annual and daily horizons. 
For the fixed window type, we set the training data to 2001-2011 and provide daily predictions for 2012-2016. In the expanding window case, we use data from 2001 up to a certain year $\tau$ in the set $\{2011, 2012, 2013, 2014\}$ and predict daily values in $\tau+l$ where $l\in \{1,2\}$\footnote{We use an expanding window here to account for the difficulty of predicting two full years at once.}. For daily rolling windows, we set the training length to $10$ years as in the initial expanding case and the fixed window setting and predict one corporate default observation ahead (Daily)\footnote{This does not necessarily mean that this is one day ahead ahead, as some defaults occurred on the same day. We chose to always jump to the next day containing a default to maintain a realistic time structure in that scenario (see also \citet{Nazemi2021}).} 
We estimate either cross-validated elastic nets (mixing parameter $\alpha=0.5$)\footnote{More specifically, the penalty in the objective function is specified as $\sum_{j=1}^p (\alpha \vert b_j \vert + (1-\alpha) b_j^2)$}, cross-validated lasso regressions, or simple linear models, and use random forests as non-parametric benchmarks.
For each window construction, we employ the post-selection methods either with the full raw data or the group-PCA transformed data (as described in Section \ref{sec:results_identification}, we use as many PCs to explain 90\% of the variance in each group, see also Table \ref{tab:selections_method_means_app}).
We either use the above data without pre-selection or employ the pre-selected set of variables according to the different setups in Section \ref{sec:results_identification}. This comprises using our proposed weighted FDR selection (wFDR)\footnote{For the wFDR, we use all PCA-components from the three most-selected groups, i.e. 14,12,11. See also Table \ref{tab:pca_lm_groups} for comparison.} combining all baseline knockoff procedures or using only our repeated-subsampling procedure (see Procedure \ref{algo:stab_sel} in Section \ref{Sec:methods}) in combination with the baseline-methods. These are either model-$X$ knockoffs (MX, or MX 2 Comp. using a maximum of 2 PCs) or deep knockoffs with 5 (Narrow) and 25 (Wide) neurons per layer.
For the infeasible reference scenario in Table \ref{tab:oos_results_full} and the group-PCA-transformed data, we use PCs that are estimated over the full data set. In the ``completely out-of-sample'' forecasting case in Table \ref{tab:oos_results_short}, the out-of-sample PCs for time points after 2011 are created using the weights from the PCs with only data up to the end of 2011\footnote{For comparability, scaling/centering uses information of the entire sample. But scaling with weights from data up to the end of 2011 does not substantially change the prediction performance. Results are available from authors upon request.}.

\begin{table}[!htb]
	\caption{Out-Of-Sample Predictions (Theoretical Infeasible Case: Model Selection and Principal Component Construction Based on the Entire Sample))} 
	\label{tab:oos_results_full}
	\centering
	\resizebox{0.9\textwidth}{!}{%
	\begin{tabular}{lcccccccc}
		\hline
		Group-PCA & Selection Method & Post-Selection  & \multicolumn{2}{c}{Fixed} & \multicolumn{2}{c}{Annual} & \multicolumn{2}{c}{Daily} \\ 
  \hline
 &  &  & RMSE & MAE & RMSE & MAE & RMSE & MAE \\ 
 \checkmark & wFDR Knock. & Elastic Net & $\mathbf{  28.74 }$ &   23.41 &   30.21 &  24.92 & $\mathbf{  29.14 }$ &  24.33 \\ 
  \checkmark & wFDR Knock. & OLS & $\mathbf{  28.50 }$ &   22.30 &   30.05 &  24.40 & $\mathbf{  29.23 }$ &  24.41 \\ 
  \checkmark & MX Knock. & Elastic Net &   28.85 & $\mathbf{  21.71 }$ & $\mathbf{  30.01 }$ & $\mathbf{ 23.61 }$ &   29.26 & $\mathbf{ 23.44 }$ \\ 
  \checkmark & MX Knock. & OLS &   28.93 & $\mathbf{  21.63 }$ & $\mathbf{  30.02 }$ & $\mathbf{ 23.52 }$ &   29.33 & $\mathbf{ 23.29 }$ \\ 
  \checkmark & MX Knock. & Random Forest &   30.87 &   26.87 &   33.19 &  28.51 &   30.87 &  25.80 \\ 
  \checkmark & MX Knock. 2 Comp. & Elastic Net &   29.86 &   24.78 &   31.50 &  26.63 &   30.09 &  25.13 \\ 
  \checkmark & MX Knock. 2 Comp. & OLS &   29.79 &   24.67 &   31.02 &  25.89 &   30.00 &  24.98 \\ 
  \checkmark & Deep Knock. Narrow & Elastic Net &   44.52 &   40.07 &   36.61 &  32.59 &   33.07 &  28.89 \\ 
  \checkmark & Deep Knock. Narrow & OLS &   43.88 &   39.46 &   36.50 &  32.49 &   33.03 &  28.83 \\ 
  \checkmark & Deep Knock. Wide & Elastic Net &   37.17 &   32.47 &   35.20 &  29.86 &   32.87 &  28.13 \\ 
  \checkmark & Deep Knock. Wide & OLS &   36.78 &   32.08 &   34.64 &  29.27 &   32.87 &  28.16 \\ 
  \checkmark & No Selection & Elastic Net &  183.94 &  162.04 &   50.25 &  37.34 &   30.87 &  23.89 \\ 
  \checkmark & No Selection & Lasso &  192.16 &  169.71 &   52.67 &  38.49 &   31.07 &  24.11 \\ 
  \checkmark & No Selection & Random Forest &   35.52 &   31.29 &   34.29 &  29.82 &   29.53 &  23.54 \\ 
   & Group Knock. & Elastic Net &  223.40 &  194.16 &   68.30 &  51.22 &   31.48 &  24.01 \\ 
   & No Selection & Elastic Net &  268.10 &  237.02 &   79.04 &  58.49 &   34.18 &  25.47 \\ 
   & No Selection & Lasso &  240.53 &  213.98 &   74.75 &  56.47 &   35.01 &  26.18 \\ 
   & No Selection & Random Forest &   36.23 &   32.03 &   33.17 &  28.99 &   29.42 &  23.55 \\  
   \hline
   \hline
	\end{tabular}
	}
	 \caption*{\footnotesize The table shows the predictive performance after different pre-selection or no pre-selection occurred, for PCA or pure data components and across different post-selection methods. In each forecasting scheme, the best two models are marked in bold. }
\end{table}

\begin{table}[!htb]
	\caption{(Completely) Out-Of-Sample Predictions (Practically Feasible Case: Model Selection and Principal Component Construction Based Only on Period up to 2012)} 
	\label{tab:oos_results_short}
	\centering
	\resizebox{0.9\textwidth}{!}{%
	\begin{tabular}{lllcccccc}
		\hline
		Group-PCA & Selection Method & Post-Selection  & \multicolumn{2}{c}{Fixed} & \multicolumn{2}{c}{Annual} & \multicolumn{2}{c}{Daily}\\ 
  \hline
 &  &  & RMSE & MAE & RMSE & MAE & RMSE & MAE \\ 
 \checkmark & wFDR Knock. & Elastic Net &   31.38 &   27.26 & $\mathbf{  31.95 }$ &  27.52 &   31.28 &  26.44 \\ 
  \checkmark & wFDR Knock. & OLS & $\mathbf{  31.12 }$ &   26.98 &   32.01 &  27.60 &   31.39 &  26.49 \\ 
  \checkmark & MX Knock. & Elastic Net &   38.67 &   32.68 &   35.18 &  29.61 &   34.80 &  29.43 \\ 
  \checkmark & MX Knock. & OLS &   38.88 &   32.83 &   35.29 &  29.70 &   34.89 &  29.49 \\ 
  \checkmark & MX Knock. & Random Forest &   31.17 & $\mathbf{  25.37 }$ &   33.15 & $\mathbf{ 26.78 }$ &   29.59 &  23.77 \\ 
  \checkmark & MX Knock. 2 Comp. & Elastic Net &   38.62 &   32.64 &   35.18 &  29.61 &   34.72 &  29.35 \\ 
  \checkmark & MX Knock. 2 Comp. & OLS &   38.88 &   32.83 &   35.29 &  29.70 &   34.89 &  29.49 \\ 
  \checkmark & No Selection & Elastic Net &  107.07 &   91.70 &   62.32 &  49.33 &   43.14 &  32.00 \\ 
  \checkmark & No Selection & Lasso &   94.93 &   80.22 &   57.54 &  46.40 &   41.31 &  31.74 \\ 
  \checkmark & No Selection & Random Forest & $\mathbf{  30.80 }$ & $\mathbf{  26.45 }$ & $\mathbf{  31.77 }$ & $\mathbf{ 27.08 }$ & $\mathbf{  28.84 }$ & $\mathbf{ 23.13 }$ \\ 
   & Group Knock. & Elastic Net &  223.40 &  194.16 &   68.30 &  51.22 &   31.48 &  24.01 \\ 
   & No Selection & Elastic Net &  268.10 &  237.02 &   79.04 &  58.49 &   34.18 &  25.47 \\ 
   & No Selection & Lasso &  240.53 &  213.98 &   70.55 &  50.16 &   34.83 &  25.81 \\ 
   & No Selection & Random Forest &   36.23 &   32.03 &   33.17 &  28.99 & $\mathbf{  29.42 }$ & $\mathbf{ 23.55 }$ \\ 
		\hline
		\hline
	\end{tabular}
	}
	 \caption*{\footnotesize It contrast to Table \ref{tab:oos_results_full}, variable selection and PCA is only performed with data up to the end of 2011. The best two models for each forecasting scheme are again marked in bold.
	 }
\end{table}

Table \ref{tab:oos_results_full} shows that generally simple linear models with limited pre-selected variables from our proposed weighted FDR procedure that combine different knockoff selections works best for forecasting both longer and shorter time horizons. Moreover, as a single selection techniques, also, the model-$X$ procedure within our robustified framework yields excellent results with a simple linear post-selection fit. Determining variables with the Deep Knockoff robustified framework generally performs slightly worse also for nonlinear post-selection models, with the relative best performance for shorter forecasting horizons. This is not unexpected since for the deep knockoffs, selection probabilities were generally much higher, meaning they could contain more noise variables that would bias predictions for longer time horizons (i.e. they do not generalize as well as the weighted FDR counterparts). The baseline linear models using the full raw data perform poorly for large time horizons, especially for fixed windows, which can be explained by potential overfitting on noisy data. Interestingly, this cannot be fully countered by regularization using elastic nets for forecasting tasks. Only for very short time horizons as the daily rolling window, the baseline procedures can compete. These findings are highly in favor of using our proposed statistical model selection techniques also for forecasting tasks. The machine learning (ML) benchmarks with selection on the full raw data perform similarly, but always slightly worse than the knockoff counterparts with the generally downside of lacking transparency and interpretability of the influence of certain groups and factors. Using the PCs instead of raw data only significantly improves the random forest model for the fixed window.
Note that for this study, we used the standard recommended data-driven choice of tuning parameters for the machine learning benchmarks but did not additionally fine-tune from there in order to maintain comparability between the simple baselines, the knockoff procedures, and the benchmark models.\footnote{While tuning all hyperparameters cautiously could improve ML-forecasts to some extent, previous studies for recovery rates show that expected changes are minor (see e.g. \citet{Nazemi2021} with additional news-based variables).}. 

In the ``completely'' out-of-sample scenario, we repeat the model selection step from Section \ref{sec:results_identification}, but only use data up to the end of 2011 to determine the relevant variables with our proposed methodology for all different knockoff-baseline procedures. This represents the most realistic but also most challenging scenario for the knockoff procedure, where post-crisis recovery rates are not contained in the training set but must be predicted. The resulting stability in selections and in forecasting performance therefore indicates that our choices are important also for non-crisis periods. Table \ref{tab:oos_results_short} summarizes the new results\footnote{We did not include the Deep Knockoff procedures here since the sample size is significantly reduced for selections.}. As expected, our proposed methodology (``wFDR") is on par with the top-performing machine learning methods in this case as well, although the random forest with the full raw data is slightly better for the daily window. In comparison to the infeasible full-sample selection results in Table \ref{tab:oos_results_full}, the single model-$X$ knockoffs perform slightly worse, with less variability between the different knockoffs employed within the subsampling knockoff framework. This can be explained by the smaller data set, where fewer variables are selected in general and difference between the different model-X knockoffs is smaller. Selections in general are the same compared to the full data case for our proposed weighted FDR procedure, and very similar for the single ``baseline" methods, with only a few minor changes (see Table \ref{tab:selections_method_prob_rank_short} in Appendix \ref{Sec:Fig_Tables} for details). In terms of forecasting performance, the ranks for the different procedures are stable, meaning that our proposed weighted FDR methodology together with the random forest are still performing best, while using no or no robust selection still performs worst overall. The margin between the latter and our procedure is smaller only for the daily rolling window and the MAE, where very bad predictions (e.g. in cases of large default events) are not punished as heavily as with the MSE. Using the usual MSE-measure, the raw data with an elastic net (or lasso) still perform considerably worse.

\begin{table}[!htb]
	\caption{Model Confidence Sets for $\alpha=0.15$ and Different Methods With Full-Sample Selection and Principal Components} 
	\label{tab:oos_results_mcs_full}
	\centering
	\resizebox{0.9\textwidth}{!}{%
	\begin{tabular}{lllcccccc}
		\hline
		& & & \multicolumn{2}{c}{Fixed} & \multicolumn{2}{c}{Annual} & \multicolumn{2}{c}{Daily} \\
		\hline
	Group-PCA & Selection Method & Post-Selection & TMax & P-Value & TMax & P-Value & TMax & P-Value \\ 
		\hline
	\multicolumn{9}{l}{\normalsize \textit{Selected Into All Model Confidence Sets}} \\
\checkmark & wFDR Knock. & Elastic Net & $\mathbf{    -5.11 }$ &      1.00 &     -0.27 &      1.00 &     -2.53 &      1.00 \\ 
  \checkmark & wFDR Knock. & OLS &     -1.36 &      1.00 &     -0.74 &      1.00 &     -2.36 &      1.00 \\ 
  \checkmark & MX Knock. & Elastic Net &     -0.78 &      1.00 & $\mathbf{    -0.99 }$ &      1.00 & $\mathbf{    -2.95 }$ &      1.00 \\ 
  \checkmark & MX Knock. & OLS &     -0.61 &      1.00 &     -0.86 &      1.00 &     -2.94 &      1.00 \\ 
  \checkmark & MX Knock. 2 Comp. & OLS &      0.88 &      0.73 &      1.70 &      0.17 &     -1.47 &      1.00 \\
   \multicolumn{9}{l}{\normalsize \textit{Selected Into Two Model Confidence Sets}} \\
  \checkmark & MX Knock. & Random Forest &      1.22 &      0.51 & - & - &     -0.43 &      1.00 \\ 
  \checkmark & MX Knock. 2 Comp. & Elastic Net &      0.95 &      0.68 & - & - &     -1.32 &      1.00 \\ 
  \multicolumn{9}{l}{\normalsize \textit{Selected Into One Model Confidence Set}} \\
 \checkmark & Deep Knock. Narrow & Elastic Net & - & - & - & - &      0.97 &      0.83 \\ 
  \checkmark & Deep Knock. Narrow & OLS & - & - & - & - &      0.94 &      0.85 \\ 
  \checkmark & Deep Knock. Wide & Elastic Net & - & - & - & - &      0.91 &      0.87 \\ 
  \checkmark & Deep Knock. Wide & OLS & - & - & - & - &      0.90 &      0.87 \\ 
  \checkmark & No Selection & Elastic Net & - & - & - & - &     -0.31 &      1.00 \\ 
  \checkmark & No Selection & Lasso & - & - & - & - &     -0.14 &      1.00 \\ 
  \checkmark & No Selection & Random Forest & - & - & - & - &     -2.12 &      1.00 \\ 
   & Group Knock. & Elastic Net & - & - & - & - &      0.19 &      1.00 \\ 
   & No Selection & Elastic Net & - & - & - & - &      1.04 &      0.80 \\ 
   & No Selection & Lasso & - & - & - & - &      1.24 &      0.66 \\ 
   & No Selection & Random Forest & - & - & - & - &     -2.37 &      1.00 \\
  \multicolumn{9}{l}{\normalsize \textit{Selected Into No Model Confidence Set}} \\
   & Group Knock. & OLS & - & - & - & - & - & - \\ 
   \hline
   \hline
	\end{tabular}
	}
	 \caption*{\footnotesize "TMax" depicts the test statistic and "P-Value" the p-value for the TMax procedure in testing equal predictive ability in the model confidence procedure of \citet{Hansen2011} implemented with the R-package from \citet{Bernardi2014}. The TMax-values depict the average loss difference of that method compared to all other methods (negative value indicating smaller loss). P-values are based on $B=5000$ bootstrap samples with rejection level set to $\alpha=0.15$. Losses are squared losses calculated as in Table \ref{tab:oos_results_full} and other details follow this table. The best model in each confidence set is marked in bold. "-" indicates that the model was not selected, i.e. had p-value smaller than $\alpha$.}
\end{table}
\begin{table}[!htb]
	\caption{Model Confidence Sets for $\alpha=0.15$ and Different Methods For the Completeley Out-Of-Sample Period With Selections up to 2012} 
	\label{tab:oos_results_mcs_short}
	\centering
	\resizebox{0.9\textwidth}{!}{%
	\begin{tabular}{lllcccccc}
		\hline
		& & & \multicolumn{2}{c}{Fixed} & \multicolumn{2}{c}{Annual} & \multicolumn{2}{c}{Daily} \\
		\hline
	Group-PCA & Selection Method & Post-Selection & TMax & P-Value & TMax & P-Value & TMax & P-Value \\ 
		\hline
   \multicolumn{9}{l}{\normalsize \textit{Selected Into All Model Confidence Sets}} \\
\checkmark & wFDR Knock. & Elastic Net &      0.40 &      0.81 &     -1.71 &      1.00 &     -1.74 &      1.00 \\ 
  \checkmark & wFDR Knock. & OLS &      0.01 &      1.00 &     -1.60 &      1.00 &     -1.66 &      1.00 \\ 
  \checkmark & MX Knock. & Random Forest &      0.06 &      1.00 &     -0.35 &      1.00 &     -3.32 &      1.00 \\ 
  \checkmark & No Selection & Random Forest & $\mathbf{    -0.61 }$ &      1.00 & $\mathbf{    -1.81 }$ &      1.00 & $\mathbf{    -4.04 }$ &      1.00 \\ 
     \multicolumn{9}{l}{\normalsize \textit{Selected Into Two Model Confidence Sets}} \\
  \checkmark & MX Knock. & Elastic Net & - & - &      1.43 &      0.34 &      0.42 &      0.99 \\ 
  \checkmark & MX Knock. & OLS & - & - &      1.44 &      0.33 &      0.48 &      0.98 \\ 
  \checkmark & MX Knock. 2 Comp. & Elastic Net & - & - &      1.43 &      0.34 &      0.37 &      0.99 \\ 
  \checkmark & MX Knock. 2 Comp. & OLS & - & - &      1.44 &      0.33 &      0.48 &      0.98 \\ 
   & No Selection & Random Forest & - & - &     -0.52 &      1.00 &     -3.32 &      1.00 \\ 
     \multicolumn{9}{l}{\normalsize \textit{Selected Into One Model Confidence Set}} \\
   \checkmark & No Selection & Elastic Net & - & - & - & - &      1.51 &      0.48 \\ 
  \checkmark & No Selection & Lasso & - & - & - & - &      1.62 &      0.40 \\ 
   & Group Knock. & Elastic Net & - & - & - & - &     -1.77 &      1.00 \\ 
   & No Selection & Elastic Net & - & - & - & - &      0.01 &      1.00 \\ 
   & No Selection & Lasso & - & - & - & - &      0.24 &      1.00 \\ 
      \multicolumn{9}{l}{\normalsize \textit{Selected Into No Model Confidence Set}} \\
   & Group Knock. & OLS & - & - & - & - & - & - \\ 
   \hline
   \hline
	\end{tabular}
	}
	 \caption*{\footnotesize Compare Table \ref{tab:oos_results_mcs_full} for detailed descriptions. Losses are squared losses calculated as in Table \ref{tab:oos_results_short} and other details follow this table.}
\end{table}

As an additional robustness check for the predictive performance of our methods, we compute model confidence sets \citep{Hansen2011} for the same forecasting combinations as before and for both the infeasible scenario and the ``pure forecasting case''. As suggested in \citet{Hansen2011}, we use $B=5000$ bootstrap replications, the \textit{TMax} test-statistic, and test-level $\alpha=0.15$. Please see Appendix \ref{sec:mcs} for details. The displayed results are robust across different $\alpha$ test levels in the standard range $[0.1; 0.2]$. \footnote{Results are omitted here but are available upon request from the authors.} Although the model confidence sets differ over the various forecasting schemes, we can identify models that consistently fall into the model confidence set. Please see Table \ref{tab:oos_results_mcs_full} and Table \ref{tab:oos_results_mcs_short} for full-sample selection and completely out-of-sample forecast results, respectively. We want to highlight that our proposed wFDR procedure always belongs to the model confidence set, together with the random forest procedure in the full-sample selection scenario. Over the full sample selection, the other model-X procedures can still compete, while in the ``completely'' out-of-sample case, they are outperformed by our proposed wFDR. This is also confirmed by the corresponding test statistics, where a lower value indicates better performance. More specifically, a negative value of the test statistic indicates that the average loss is smaller compared to all other methods in the confidence set, where our suggested method clearly outperforms the other procedures in general for the full-sample selection, while for the ``completely'' out-of-sample scenario, the raw random forest is slightly better when comparing the test-statistic. Using no or non-robust selection methods (i.e. lasso, elastic net) is always worse, and also the plain group knockoff does not perform well on our data set, which might be caused by our specific data structure, where we still have some strong correlations between groups.

\section{Conclusion} \label{Sec:Conclusion}
In this paper, we demonstrate the benefit of connecting the flexibility of the knockoff framework with repeated subsampling and techniques controlling the proportion of false discoveries over the full spectrum of possible values. We employ a comprehensive set of distinct knockoff machines and illustrate that a transparent combination of their results yields optimal ensemble results.

With the proposed methodology, we are able to uncover important macroeconomic factors of corporate bond recovery rates while maintaining excellent forecasting performance. In particular, predictive power in various settings using linear models with just the identified groups is significantly higher than using the full set of variables in similar models. Furthermore, our procedure outperforms other model selection procedures and performs similar to flexible machine learning methods. The latter are developed for prediction tasks but lack easy interpretation and identification of important factors in contrast to the proposed methodology.

For future research, the proposed technique shows high-potential in other data-rich environments, such as asset pricing or climate modelling. In a separate paper, it would be of interest to derive conditions for FDR-level components and optimized forms of parsimonious weighting schemes to theoretically achieve and derive optimal in-sample or out-of-sample fits and respective statistical rates.

\newpage
\bibliography{References}


\newpage

\appendix

\section{Methods}
\subsection{Deep Knockoffs} \label{sec:Additional_Methods}
Even though the procedure of \citet{Candes2018} poses only few assumptions on $Y$ and $X$ together, namely that the observations are identically and independently distributed, there is the assumption that the distribution $F_{X}$ of $X$ is known beforehand. This procedure can be ineffective in producing reliable knockoff variables when the covariance structure of $X$ is hard to replicate, e.g. when variables in $X$ are highly correlated. This problem arises due to often conflicting requirements of $X$ and $\Tilde{X}$ to have a similar covariance structure but to be uncorrelated at the same time. \citet{Romano2019} propose to solve these issues by replacing the model-$X$ algorithm by an artificial neural network. They define the covariance matrix of the combined $(X,\Tilde{X})$ $G$ and t:
\begin{align}
	G=Cov[(X,\Tilde{X})]=
	\begin{bmatrix}
		G_{XX} & G_{X\Tilde{X}}\\
		G_{\Tilde{X}X} & G_{\Tilde{X}\Tilde{X}}
	\end{bmatrix} \ .
\end{align}

We apply this approach to improve the creation of knockoffs and to provide a robustness check of the model-$X$ knockoffs. In contrast to the model-$X$ construction, a deep neural network is used to generate knockoff variables. We use different structures of neural networks and compare their performance. In the final analysis, we include one wide network (25 neurons per layer) and one narrow network (5 neurons per layer), each of them with six layers, and otherwise the same structure as in the implementation of \citet{Romano2019}\footnote{See \url{https://github.com/msesia/deepknockoffs} for details.}. The advantage of the Deep Knockoff procedure lies in the creation of the knockoffs. Since we use neural networks, we can model more complex relations and control specifically for higher moments in the knockoff distribution as well as the correlation of $X$ and $\Tilde{X}$.

Given $X$ and a random noise matrix $V$, the network outputs knockoff copies $\Tilde{X}$ that are then evaluated using a customized loss function. Also define $X'$, $X'' \in \mathbb{R}^{n/2 \times p}$ as a random partition of $X$. This loss function $J$ can be defined as follows (see \citet{Romano2019} for details) given $M$ as a $p\times p$-matrix with zero diagonal and ones everywhere else, $\circ$ as element-wise multiplication of two matrices: 
\begin{align*}
	J_{\gamma,\lambda,\delta}(X,\Tilde{X})&= \gamma J_{MMD}(X,\Tilde{X}) +\lambda J_{second-order}(X,\Tilde{X}) +\delta J_{decorrelation}(X,\Tilde{X}) \\
	J_{MMD}(X,\Tilde{X})&= \hat{D}_{MMD}\left[(X',\Tilde{X}'),(\Tilde{X}'',X'')\right] +\hat{D}_{MMD}\left[(X',\Tilde{X}'),(X'',\Tilde{X}'')_{swap(S)}\right] \\
	J_{second-order}(X,\Tilde{X})&= \lambda_1 \dfrac{\Vert G_{XX} - G_{\Tilde{X}\Tilde{X}}\Vert^2_2}{\Vert G_{XX}\Vert^2_2} + \lambda_2 \dfrac{\Vert M \circ (G_{XX} - G_{\Tilde{X}\Tilde{X}})\Vert^2_2}{\Vert G_{XX}\Vert^2_2} + \\ 
	&\dfrac{\lambda_3}{p} \left\Vert \dfrac{ \sum_{i=1}^n (X_i-\Tilde{X}_i)}{n}\right\Vert^2_2 \\
	J_{decorrelation-order}(X,\Tilde{X})&= \Vert \text{diag}(G_{X\Tilde{X}}) -1 + s_{ASDP}^*(G_{XX}) \Vert^2_2  \ .
\end{align*}
$\lambda=(\lambda_1,\lambda_2,\lambda_3)$, $s_{ASDP}^*(\Omega)$ is a function returning the optimal $s^*=(s_1^*,\dots,s_p^*)$ from the ASDP-procedure for model-$X$ knockoffs given a covariance matrix $\Omega$. $\hat{D}_{MMD}(X,Z\Tilde{X})$ is the empirical version of the maximum mean discrepancy using a Gaussian kernel for comparing two matrices $X$ and $\Tilde{X}$. $(X,\Tilde{X})_{swap(S)}$ stands for matrix $(X,\Tilde{X})$ with entries of $X$ and $\Tilde{X}$ swapped in dimensions $S\in \{1,\dots,p\}$, where each dimension $j$ is contained in $S$ with probability 0.5. This loss function contains three parts. $J_{second-order}(X,\Tilde{X})$ measures the deviation from the first moment in $\lambda_3$, and the deviation from the diagonal ($\lambda_1$) as well as the off-diagonal elements ($\lambda_2$) in $G$. $J_{MMD}(X,\Tilde{X})$ penalizes discrepancies in the two covariate-distributions in general, i.e. targeting higher moments. This is done in computationally efficient way by computing the MMD-distance on differently arranged versions of the two independent samples $X'$, $X''$. Finally, $J_{decorrelation}(X,\Tilde{X})$ is added to ensure that the knockoffs $\Tilde{X}$ are not highly correlated with $X$. Otherwise, the algorithm could easily find an optimal trivial solution in just setting $X=\Tilde{X}$. The degrees of influence of each of these parts are set by $\gamma$, $\lambda$, and $\delta$, which should be set depending on the underlying data.

\subsection{Group Knockoffs}
We additionally employ the group knockoff filter from \citet{Dai2016} as a robustness-check, since it is supposed to handle highly correlated variables better by imposing a group structure. The selection step using the lasso-signed max statistic is taken over a group-lasso regression, encouraging group-sparsity in the selection. $W_j^{LSM}$ is changed accordingly so that groups of variables can are selected in the end. This is simply done by replacing the individual coefficients by their group counterparts and thus recording at which $\lambda$ those are included into the model. The construction of good knockoffs, however, is easier due to the imposed structure, and is an extension to what is done in \citet{Barber2015}. Intuitively, variables that are highly correlated should be in the same group, and the procedure should work well correlations among groups are low. In our case, this is not fully fulfilled, which is why we expect the performance to be lower than for the other methods. 

\subsection{Random Forest} \label{sec:Random_forest}
We also use random forests \citep{Breiman2001} both for prediction as a fully non-parametric machine learning benchmark with a tree-size of 2000. We can extract variable importance from this procedure based on mean variance reduction as a robustness check. This measures the mean reduction in mean squared error by splitting on a certain variable. To aggregate this on a group level, we take the mean over the reduction of all the variables in the respective group. We additionally extract p-values using the PIMP-procedure suggested in \citet{Altmann2010}. There, we use 200 permutations of the response variable and measure the variable importances for each permutation to obtain 200 base-importances for each variable. We can then fit a distribution to these \textit{null} importances to obtain a null distribution against which we compare the extracted true variable importance to obtain the p-value. We use a simple and fast nonparametric approach to obtain the p-values by simply measuring the fraction of null importances that exceed the true measured importance relative to the number of permutations. With that, we follow the suggestion of \citet{Altmann2010} and the subsequent implementation in the \texttt{R}-package \texttt{ranger}.

\subsection{Model Confidence Sets} \label{sec:mcs}
While our focus is on interpretation of data-driven selected model components, we also study the predictive ability of the resulting knock-off determined models. In this setting, we compute model confidence sets proposed in \citet{Hansen2011} and implemented in \citet{Bernardi2014}. Such model confidence sets provide practitioners with more robust statistical guidance on which models to apply rather then using simple prediction errors. Intuitively, the procedure proceeds iteratively, testing whether all models have the same predictive ability, eliminating the worst model, until equal predictive ability cannot be rejected. Define the loss series as in \citet{Bernardi2014}, i.e $d_{ijt}=l_{it}-l_{jt}$,$i,j \in M=\{1,\dots,m\}$, as the difference of the squared loss $l_{it}$ and $l_{jt}$ at time point $t=1,\dots,T$ for model $i$ and $j$ out of $m$ available models. To construct the test statistic, we compute $d_{i\cdot t}=(m-1)^{-1} \sum_{j\in M\setminus i} d_{ijt}$, the average loss difference between model $i$ and all other (remaining) models. We then construct the statistic $t_{i\cdot}=\dfrac{\overline{d}_{i\cdot}}{\sqrt{\widehat{var}(\overline{d_{i\cdot}})}}$, where $\overline{d_{i\cdot}}=\dfrac{1}{T}\sum_{t=1}^T d_{i\cdot t}$. $\widehat{var}(\overline{d_{i\cdot}})$ is the bootstrapped variance using the block-bootstrap with $B=5000$ bootstrap samples and a block length $k$ that is equal to the number of selected parameters in an auto-regression of the the loss difference series $d_{ijt}$ using the Akaike information criterion to select the model order. The test statistic we use is $T_{max,M}=\underset{i\in M}{max} \ t_{i\cdot}$, and the test rejects if this value is larger than the $1-\alpha$ quantile of the bootstrapped distribution of $T_{max,M}$.

\begin{table}[ht]
\centering
\caption{Most-Selected Groups Over Different Weighting Schemes For Completely Out-Of-Sample Selections} 
\label{tab:selections_method_prob_rank_short}
\resizebox{0.9\textwidth}{!}{%
\begin{tabular}{lrrrrrrrr}
  \hline
  Rank: & \multicolumn{2}{c}{1} & \multicolumn{2}{c}{2} & \multicolumn{2}{c}{3} & \multicolumn{2}{c}{4} \\
  \hline
 & Group & Mean Score & Group & Mean Score & Group & Mean Score & Group & Mean Score \\ 
  \hline
Prob\_unweight & 14 & 0.648 & 12 & 0.631 & 11 & 0.596 & 5 & 0.534 \\ 
  Prob\_exp & 5 & 0.407 & 12 & 0.402 & 14 & 0.376 & 20 & 0.354 \\ 
  Prob\_lin & 12 & 0.506 & 14 & 0.493 & 5 & 0.462 & 11 & 0.458 \\ 
  Rank\_unweight & 14 & 17.460 & 12 & 15.747 & 11 & 15.032 & 15 & 14.636 \\ 
  Rank\_exp & 20 & 16.876 & 14 & 16.495 & 12 & 15.919 & 11 & 14.853 \\ 
  Rank\_lin & 14 & 16.990 & 20 & 16.014 & 12 & 15.779 & 11 & 14.951 \\ 
   \hline
\end{tabular}
}
 \caption*{\footnotesize Note: In the columns, \textit{Group} depicts the selected variable group, while \textit{Mean Score} shows the (weighted) mean over all five procedures for the four most selected groups. The ranks for groups are rescaled linearly to lie between 1 and 20 (20 being the best score), while the selection probabilities lie between 0 and 1.} 
\end{table}


	\center
\begin{sidewaystable}[htbp]
  \centering
  \caption{Groups of independent variables}
	\tiny
    \resizebox*{0.9\textwidth}{!}{%
    \begin{tabular}{ll}
				\hline
    \textbf{Group 1: Financial Conditions: Loans} &  \\
		\hline
      Total Net Loan Charge-offs to Total Loans for Banks & Nonperforming Total Loans (past due 90+ days plus nonaccrual) to Total Loans \\
    Nonperforming Loans to Total Loans (avg assets betw. USD 100M and 300M) & Net Loan Losses to Average Total Loans for all U.S. Banks \\
    Loan Loss Reserve to Total Loans for all U.S. Banks & Nonperforming Commercial Loans (past due 90+ days plus nonaccrual) to Commercial Loans \\
		\hline
    \textbf{Group2: Monetary Measures: Savings} &  \\
		\hline
     Personal Saving & Personal Saving Rate \\
    Gross Saving &  \\
		\hline
    \textbf{Group 3: Monetary Measures: CPIs} &  \\
		\hline
    Gross Domestic Product: Implicit Price Deflator & Consumer Price Index for All Urban Consumers: All Items Less Food \\
    University of Michigan Inflation Expectation & Consumer Price Index for All Urban Consumers: Energy \\
    Consumer Price Index for All Urban Consumers: Apparel & Consumer Price Index for All Urban Consumers: All Items \\
    Consumer Price Index for All Urban Consumers: Medical Care & Consumer Price Index for All Urban Consumers: Transportation \\
    Consumer Price Index for All Urban Consumers: All items less shelter & Consumer Price Index for All Urban Consumers: All items less medical care \\
    Consumer Price Index for All Urban Consumers: Durables & Consumer Price Index for All Urban Consumers: Services \\
    Consumer Price Index for All Urban Consumers: Commodities &  \\
    	\hline
    \textbf{Group 4: Monetary Measures: Money Supply} &  \\
		\hline
    M2 Money Stock & Board of Governors Monetary Base, Adjusted for Changes in Reserve Requirements \\
    M1 Money Stock & M3 for the United States \\
    
    		\hline
    \textbf{Group 5: Corporate Measures: Cash Flow and Profit} &  \\
		\hline
    Corporate Profits After Tax (without IVA and CCAdj) & Corporate Profits After Tax with Inventory Valuation and Capital Consumption Adjustments \\
    Corporate Profits after tax with IVA and CCAdj: Net Dividends & Corporate Net Cash Flow with IVA \\
    
		\hline
    \textbf{Group 6: Business Cycle: Unemployment} &  \\
		\hline
   Initial Unemployment Claims & \multicolumn{1}{l}{Persons unemployed 15 weeks or longer, as a percent of the civilian labor force} \\
   Civilian Unemployment Rate & \multicolumn{1}{l}{Continued Claims (Insured Unemployment)} \\
   Number of Civilians Unemployed for 5 to 14 Weeks & \multicolumn{1}{l}{Number of Civilians Unemployed for 15 Weeks and Over} \\
    Number of Civilians Unemployed for 15 to 26 Weeks & \multicolumn{1}{l}{Number of Civilians Unemployed for 27 Weeks and Over} \\
    Number of Civilians Unemployed for Less Than 5 Weeks & \multicolumn{1}{l}{Average (Mean) Duration of Unemployment} \\

    	\hline
    \textbf{Group 7: Business Cycle: Industrial Production} &  \\
		\hline
    Industrial Production Index & \multicolumn{1}{l}{University of Michigan: Consumer Sentiment} \\
    Industrial Production: Business Equipment & \multicolumn{1}{l}{Industrial Production: Consumer Goods} \\
    Industrial Production: Durable Consumer Goods & \multicolumn{1}{l}{Industrial Production: Durable Materials} \\
    Industrial Production: Final Products (Market Group) & \multicolumn{1}{l}{Industrial Production: Fuels} \\
    Industrial Production: Manufac
    turing (SIC) & \multicolumn{1}{l}{Industrial Production: Materials} \\
    Industrial Production: Nondurable Consumer Goods & \multicolumn{1}{l}{Industrial Production: Nondurable Materials} \\
    Industrial Production: Manufacturing (NAICS) &  \\

    \end{tabular}%
    }
  \label{tab:groups}%
\end{sidewaystable}%
	\center

\addtocounter{table}{-1}
\begin{sidewaystable}[htbp]
  \centering
  \caption{(Continued)}
	\tiny
    \resizebox*{0.9\textwidth}{!}{%
    \begin{tabular}{ll}

    \hline
    \textbf{Group 8: Business Cycle: Private Employment} &  \\
		\hline
    
    Average Weekly Hours of Production and Nonsupervisory Employees: Mfg & \multicolumn{1}{l}{Civilian Employment} \\
    Civilian Employment-Population Ratio & \multicolumn{1}{l}{Nonfarm Private Construction Payroll Employment} \\
    Nonfarm Private Financial Activities Payroll Employment & \multicolumn{1}{l}{Nonfarm Private Goods - Producing Payroll Employment} \\
    Nonfarm Private Manufacturing Payroll Employment & \multicolumn{1}{l}{Nonfarm Private Service - Providing Payroll Employment} \\
    Total Nonfarm Private Payroll Employment & \multicolumn{1}{l}{Nonfarm Private Trade, Transportation, and Utilities Payroll Employment} \\

    \hline
    \textbf{Group 9: Business Cycle: Housing Market} &  \\
		\hline
    New Private Housing Units Authorized by Building Permits & \multicolumn{1}{l}{Housing Starts: Total: New Privately Owned Housing Units Started} \\
    Housing Starts: Total: New Privately Owned Housing Units Started & \multicolumn{1}{l}{New One Family Houses Sold: United States} \\
    Housing Starts in Midwest Census Region & \multicolumn{1}{l}{Housing Starts in Northeast Census Region} \\
    Housing Starts in South Census Region & \multicolumn{1}{l}{Housing Starts in West Census Region} \\
    New Private Housing Units Authorized by Building Permits in the Midwest & \multicolumn{1}{l}{New Private Housing Units Authorized by Building Permits in the Northeast} \\
    New Private Housing Units Authorized by Building Permits in the South & \multicolumn{1}{l}{New Private Housing Units Authorized by Building Permits in the West} \\

    	\hline
    \textbf{Group 10: Business Cycle: Income} &  \\
		\hline
    Growth rate of Nominal Dispoable Income & \multicolumn{1}{l}{Real Disposable Personal Income} \\
    National income & \multicolumn{1}{l}{Personal Income} \\
   
    			\hline
    \textbf{Group 11: Stock Market: Index Returns and Volatilities} &  \\
		\hline
    S\&P 500 Index return & S\&P 500 Volatility 1m  \\
    CBOE DJIA Volatility Index & NASDAQ 100 Index return \\
    CBOE NASDAQ 100 Volatility Index & Russell 2000 Price Index return \\
    Russell 2000 Vol 1m & Wilshire US Small-Cap Price Index return \\
    Wilshire Small Cap Vol &  \\
		\hline
    \textbf{Group 12: International Competitiveness: Exchange Rates} &  \\
		\hline
    Canada / U.S. Foreign Exchange Rate, Canadian Dollars to One U.S. Dollar & Japan / U.S. Foreign Exchange Rate, Japanese Yen to One U.S. Dollar \\
    Switzerland / U.S. Foreign Exchange Rate, Swiss Francs to One U.S. Dollar & U.S. / U.K. Foreign Exchange Rate, U.S. Dollars to One British Pound \\
    Real Broad Effective Exchange Rate for United States &  \\

    		\hline
    \textbf{Group 13: International Competitiveness: Trade} &  \\
		\hline
    Real Trade Weighted U.S. Dollar Index: Broad & Trade Weighted U.S. Dollar Index: Major Currencies \\
    Total Current Account Balance for the United States & Real Exports of Goods \& Services \\
    Real imports of goods and services & \\

    \end{tabular}%
    }
\end{sidewaystable}%

	\center
\addtocounter{table}{-1}

\begin{sidewaystable}[htbp]
  \centering
  \caption{(Continued)}
	\tiny
	\resizebox*{0.9\textwidth}{!}{%
    \begin{tabular}{ll}
    
    	\hline
    \textbf{Group 14: Micro-level: Bond Yields and Interest Rates} &  \\
		\hline
    Bank Prime Loan Rate & 1-Month AA Nonfinancial Commercial Paper Rate \\
    10-Year Treasury Constant Maturity Rate & 3-Month AA Nonfinancial Commercial Paper Rate \\
    Term Structure & Effective Federal Funds Rate \\
    Moody's Seasoned Baa Corporate Yield Relative to Yield on 10-Year Treasury  & Moody's Seasoned Aaa Corporate Bond Yield \\
    30-Year Conventional Mortgage Rate & Moody's Seasoned Baa Corporate Bond Yield \\
    1-Year Treasury Constant Maturity Rate & 5-Year Treasury Constant Maturity Rate \\
    3-Month Treasury Bill: Secondary Market Rate & 3-month Treasury Constant Maturity Rate \\
    6-Month Treasury Bill: Secondary Market Rate & Moody's Seasoned Aaa Corporate Bond Minus Federal Funds Rate \\
    Moody's Seasoned Baa Corporate Bond Minus Federal Funds Rate & 3-Month Commercial Paper Minus Federal Funds Rate \\
    Moody's Seasoned Aaa Bbb Spread & Size of High Yield Market in U.S. Dollars \\

    		\hline
    \textbf{Group 15: Micro-level: Bond Defaults in Industry} &  \\
		\hline
   
    Bond defaults within the industry (in percent) &  \\
    
    		\hline
    \textbf{Group 16: Micro-level: High Yield Default Rate} &  \\
		\hline
    High Yield Default Rate, Trailing 12-month & \\
    
    	\hline
    \textbf{Group 17: Financial Conditions: Bank Credit and Debt} &  \\
		\hline
    Loans and Leases in Bank Credit, All Commercial Banks & Real Estate Loans, All Commercial Banks \\
    Federal Debt: Total Public Debt & Total Consumer Credit Owned and Securitized, Outstanding \\
    Excess Reserves of Depository Institutions & Commercial and Industrial Loans, All Commercial Banks \\
    Total Borrowings of Depository Institutions from the Federal Reserve & Bank Credit of All Commercial Banks \\
    Household Debt Service Payments as a Percent of Disposable Personal Income & Household Financial Obligations as a percent of Disposable Personal Income \\
    Loans and Leases in Bank Credit, All Commercial Banks &  \\
    
    	\hline
    \textbf{Group 18: Business Cycle: Real GDP} &  \\
		\hline
    Real Gross Domestic Product & \multicolumn{1}{l}{Government Consumption Expenditures} \\
    Growth rate of nominal GDP &  \\

		\hline
    \textbf{Group 19: Micro-level: Producer Price Index} &  \\
		\hline

    Producer Price Index by Commodity Industrial Commodities & Producer Price Index by Commodity Intermediate Energy Goods \\
    Producer Price Index by Commodity for Crude Energy Materials & Producer Price Index by Commodity for Finished Consumer Goods \\
    Producer Price Index by Commodity Intermediate Materials & Producer Price Index for All Commodities \\

    	\hline
    \textbf{Group 20: Business Cycle: Inventories} &  \\
		\hline
    Capacity Utilization: Manufacturing & \multicolumn{1}{l}{Change in Private Inventories} \\
    Capacity Utilization: Total Industry & \multicolumn{1}{l}{Total Business Inventories} \\
    
    \end{tabular}%
    }
\end{sidewaystable}%


\begin{table}[ht]
	\centering
	\caption{Variable Importance from Random Forests Aggregated on Group Level}
	\begin{tabular}{rrr}
		\hline
		& Variance Reduction & P-Value \\ 
		\hline
		  Group\_15 & 4.533 & 0.005 \\ 
          Group\_12 & 2.411 & 0.005 \\ 
          Group\_11 & 1.678 & 0.008 \\ 
          Group\_19 & 1.481 & 0.005 \\ 
          Group\_2 & 1.433 & 0.085 \\ 
          Group\_14 & 1.162 & 0.008 \\ 
          Group\_13 & 1.046 & 0.027 \\ 
          Group\_17 & 0.902 & 0.046 \\ 
          Group\_3 & 0.870 & 0.016 \\ 
          Group\_20 & 0.763 & 0.091 \\ 
          Group\_4 & 0.602 & 0.007 \\ 
          Group\_8 & 0.574 & 0.020 \\ 
          Group\_6 & 0.535 & 0.008 \\ 
          Group\_7 & 0.529 & 0.071 \\ 
          Group\_5 & 0.393 & 0.139 \\ 
          Group\_1 & 0.321 & 0.112 \\ 
          Group\_9 & 0.298 & 0.140 \\ 
          Group\_10 & 0.247 & 0.128 \\ 
          Group\_18 & 0.138 & 0.320 \\ 
          Group\_16 & 0.085 & 0.075 \\ 
		\hline
	\end{tabular}
	 \caption*{\footnotesize Note: Variance reduction is the mean variance reduction (i.e. influence) of a variable in the random forest, measured by the mean reduction in mean squared error by splitting on this variable, averaged over all groups. For ease of presentation, we show values relative to the average of variance reduction taken over all variables (i.e. a value larger than 1 indicates higher importance). P-Value depicts the mean p-values obtained by the PIMP-procedure from \cite{Altmann2010} given 200 permutations.}
	\label{tab:rf_altmann_importance}
\end{table}

\begin{table}[ht]
\centering
\caption{Mean Selection Probabilities for Each Procedure over all Weighting Schemes} 
\label{tab:selections_method_means_app}
\resizebox*{!}{0.8\textheight}{%
\begin{tabular}{lrrrrlr}
  \hline
Method: & modelX\_PCA & deep5\_PCA & deep25\_PCA & modelX2comp\_PCA & Group & gKnock\_Data \\ 
  \hline
PC1.1 & 0.125 & 0.695 & 0.357 & 0.212 & G\_1 & 0.081 \\ 
  PC1.2 & 0.884 & 0.990 & 0.980 & 0.279 & G\_2 & 0.313 \\ 
  PC2.1 & 0.298 & 0.990 & 0.899 & 0.227 & G\_3 & 0.195 \\ 
  PC2.2 & 0.116 & 0.952 & 0.561 & 0.137 & G\_4 & 0.099 \\ 
  PC3.1 & 0.103 & 0.526 & 0.354 & 0.085 & G\_5 & 0.080 \\ 
  PC3.2 & 0.129 & 0.683 & 0.336 & 0.105 & G\_6 & 0.073 \\ 
  PC4.1 & 0.125 & 0.677 & 0.239 & 0.087 & G\_7 & 0.148 \\ 
  PC5.1 & 0.926 & 0.990 & 0.980 & 0.587 & G\_8 & 0.079 \\ 
  PC5.2 & 0.170 & 0.888 & 0.591 & 0.116 & G\_9 & 0.179 \\ 
  PC6.1 & 0.107 & 0.606 & 0.350 & 0.084 & G\_10 & 0.157 \\ 
  PC6.2 & 0.193 & 0.950 & 0.551 & 0.129 & G\_11 & 0.202 \\ 
  PC7.1 & 0.159 & 0.990 & 0.852 & 0.103 & G\_12 & 0.264 \\ 
  PC7.2 & 0.259 & 0.990 & 0.970 & 0.384 & G\_13 & 0.112 \\ 
  PC7.3 & 0.361 & 0.990 & 0.980 &  & G\_14 & 0.294 \\ 
  PC8.1 & 0.113 & 0.662 & 0.317 & 0.105 & G\_15 & 0.464 \\ 
  PC8.2 & 0.115 & 0.958 & 0.612 & 0.085 & G\_16 & 0.120 \\ 
  PC9.1 & 0.294 & 0.990 & 0.905 & 0.131 & G\_17 & 0.098 \\ 
  PC10.1 & 0.109 & 0.468 & 0.194 & 0.089 & G\_18 & 0.321 \\ 
  PC10.2 & 0.227 & 0.990 & 0.846 & 0.200 & G\_19 & 0.256 \\ 
  PC11.1 & 0.065 & 0.985 & 0.750 & 0.004 & G\_20 & 0.388 \\ 
  PC11.2 & 0.709 & 0.990 & 0.977 & 0.763 &  &  \\ 
  PC11.3 & 0.310 & 0.990 & 0.871 &  &  &  \\ 
  PC11.4 & 0.262 & 0.990 & 0.938 &  &  &  \\ 
  PC12.1 & 0.141 & 0.990 & 0.889 & 0.215 &  &  \\ 
  PC12.2 & 0.721 & 0.990 & 0.977 & 0.616 &  &  \\ 
  PC13.1 & 0.111 & 0.722 & 0.386 & 0.093 &  &  \\ 
  PC13.2 & 0.184 & 0.990 & 0.830 & 0.100 &  &  \\ 
  PC13.3 & 0.306 & 0.990 & 0.976 &  &  &  \\ 
  PC14.1 & 0.489 & 0.990 & 0.843 & 0.116 &  &  \\ 
  PC14.2 & 0.793 & 0.990 & 0.980 & 0.848 &  &  \\ 
  PC14.3 & 0.025 & 0.688 & 0.376 &  &  &  \\ 
  PC15.1 & 0.361 & 0.990 & 0.783 & 0.422 &  &  \\ 
  PC16.1 & 0.307 & 0.990 & 0.840 & 0.496 &  &  \\ 
  PC17.1 & 0.109 & 0.535 & 0.268 & 0.084 &  &  \\ 
  PC17.2 & 0.127 & 0.609 & 0.266 & 0.125 &  &  \\ 
  PC17.3 & 0.119 & 0.880 & 0.376 &  &  &  \\ 
  PC18.1 & 0.111 & 0.598 & 0.389 & 0.389 &  &  \\ 
  PC18.2 & 0.208 & 0.990 & 0.905 & 0.264 &  &  \\ 
  PC19.1 & 0.120 & 0.601 & 0.267 & 0.086 &  &  \\ 
  PC20.1 & 0.372 & 0.990 & 0.980 & 0.182 &  &  \\ 
  PC20.2 & 0.118 & 0.769 & 0.384 & 0.130 &  &  \\ 
   \hline
\end{tabular}
  }
 \caption*{\footnotesize Note: Mean selection probabilities over the three weighting schemes for each procedure and variable. For ranking the variables, e.g. in the forecasting, the higher index is chosen first in case two probabilities are exactly the same (only relevant for the deep knockoff procedures).  }
\end{table}

\begin{table}[ht]
\centering
\caption{PCA-Weights for Groups of our Proposed Procedure} 
\label{tab:pca_weights}
\resizebox*{!}{0.9\textheight}{%
\begin{tabular}{lrrrr}
  \hline
 & PC1 & PC2 & PC3 & PC4 \\ 
  \hline
 \textit{Group\_14} \\[1.8ex]

  DCPN30 & -0.267 & 0.046 & 0.088 & - \\ 
  DGS10 & -0.217 & -0.244 & -0.22 & - \\ 
  DCPN3M & -0.267 & 0.043 & 0.108 & - \\ 
  TermStructure & -0.267 & 0.043 & -0.008 & - \\ 
  FEDFUNDS & -0.266 & 0.033 & 0.059 & - \\ 
  BAA10YM & 0.117 & -0.27 & 0.463 & - \\ 
  DAAA & -0.176 & -0.37 & -0.109 & - \\ 
  MORTGAGE30US & -0.222 & -0.265 & -0.009 & - \\ 
  DBAA & -0.103 & -0.433 & 0.249 & - \\ 
  DPRIME & -0.266 & 0.052 & 0.072 & - \\ 
  DGS1 & -0.268 & 0.016 & 0.061 & - \\ 
  DGS5 & -0.244 & -0.16 & -0.161 & - \\ 
  DTB6 & -0.268 & 0.039 & 0.061 & - \\ 
  TB3MS & -0.268 & 0.027 & 0.037 & - \\ 
  AAAFF & 0.203 & -0.313 & -0.157 & - \\ 
  BAAFF & 0.194 & -0.343 & 0.104 & - \\ 
  CPFF & 0.002 & -0.05 & 0.71 & - \\ 
  AaaBbbSpread & 0.203 & -0.313 & -0.157 & - \\ 
  HYMSIZE & 0.165 & 0.342 & 0.192 & - \\ 
  DGS3MO & -0.267 & 0.043 & -0.008 & - \\[1.8ex] 
  \textit{Group\_11} \\[1.8ex]
  SPRet & 0.427 & 0.189 & -0.053 & 0.053 \\ 
  VolSP500 & -0.351 & 0.365 & -0.292 & -0.191 \\ 
  VXDCLS & -0.374 & 0.31 & -0.258 & -0.171 \\ 
  NasdaqRet & 0.357 & 0.352 & -0.208 & -0.391 \\ 
  VXNCLS & -0.331 & 0.211 & 0.21 & 0.664 \\ 
  RussellRet & 0.38 & 0.357 & 0.097 & 0.1 \\ 
  Russell2000Vol1m & 0.154 & -0.097 & -0.833 & 0.479 \\ 
  WilshireRet & 0.384 & 0.05 & 0.173 & 0.266 \\ 
  WilshireVol1m & 0.049 & -0.654 & -0.155 & -0.164 \\[1.8ex] 
  \textit{Group\_12} \\[1.8ex]
  DEXCAUS & -0.51 & 0.131 & - & - \\ 
  DEXJPUS & -0.4 & -0.561 & - & - \\ 
  DEXSZUS & -0.476 & -0.222 & - & - \\ 
  DEXUSUK & 0.305 & -0.785 & - & - \\ 
  RBUSBIS & -0.51 & 0.047 & - & - \\[1.8ex] 
  \textit{Group\_20} \\[1.8ex]
  CAPUTLB00004SQ & -0.568 & 0.134 & - & - \\ 
  CBI & -0.531 & 0.176 & - & - \\ 
  TCU & -0.564 & 0.173 & - & - \\ 
  BUSINV & -0.278 & -0.96 & - & - \\[1.8ex]
  \textit{Group\_5} \\[1.8ex]
  CP & 0.54 & -0.072 & - & - \\ 
  CPATAX & 0.545 & 0.193 & - & - \\ 
  DIVIDEND & 0.418 & -0.805 & - & - \\ 
  CNCF & 0.487 & 0.556 & - & - \\ 
   \hline
\end{tabular}
}
\end{table}
	

	\begin{figure}
		\centering
		\includegraphics[width=0.5\textwidth]{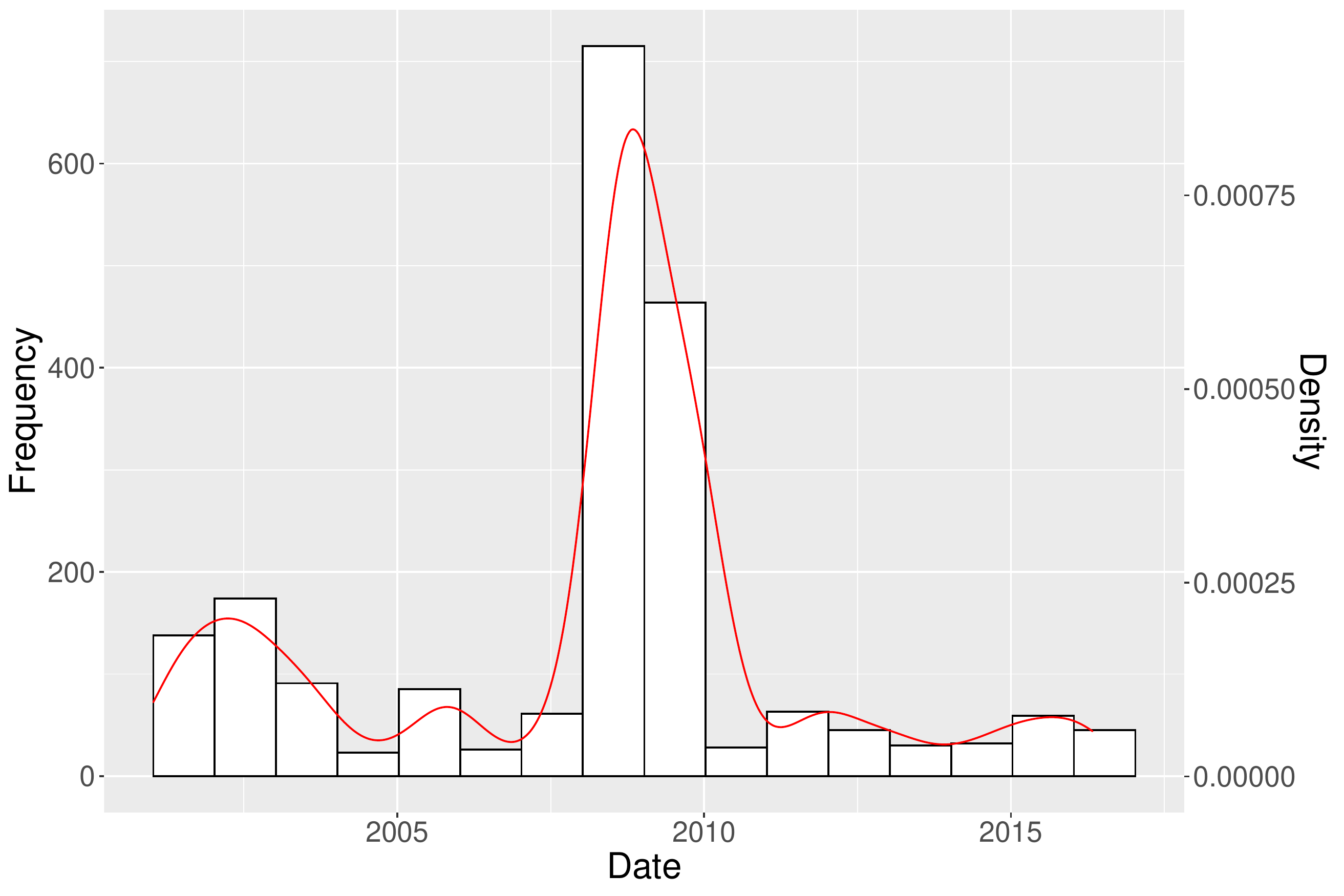}
		\caption{Default frequency and density (red) over time for the defaulted US corporate bonds from 2001 to 2016.}
		\label{fig:Defaults_hist_plot}
	\end{figure}
	
	\begin{figure}
		\centering
		\includegraphics[width=0.7\textwidth]{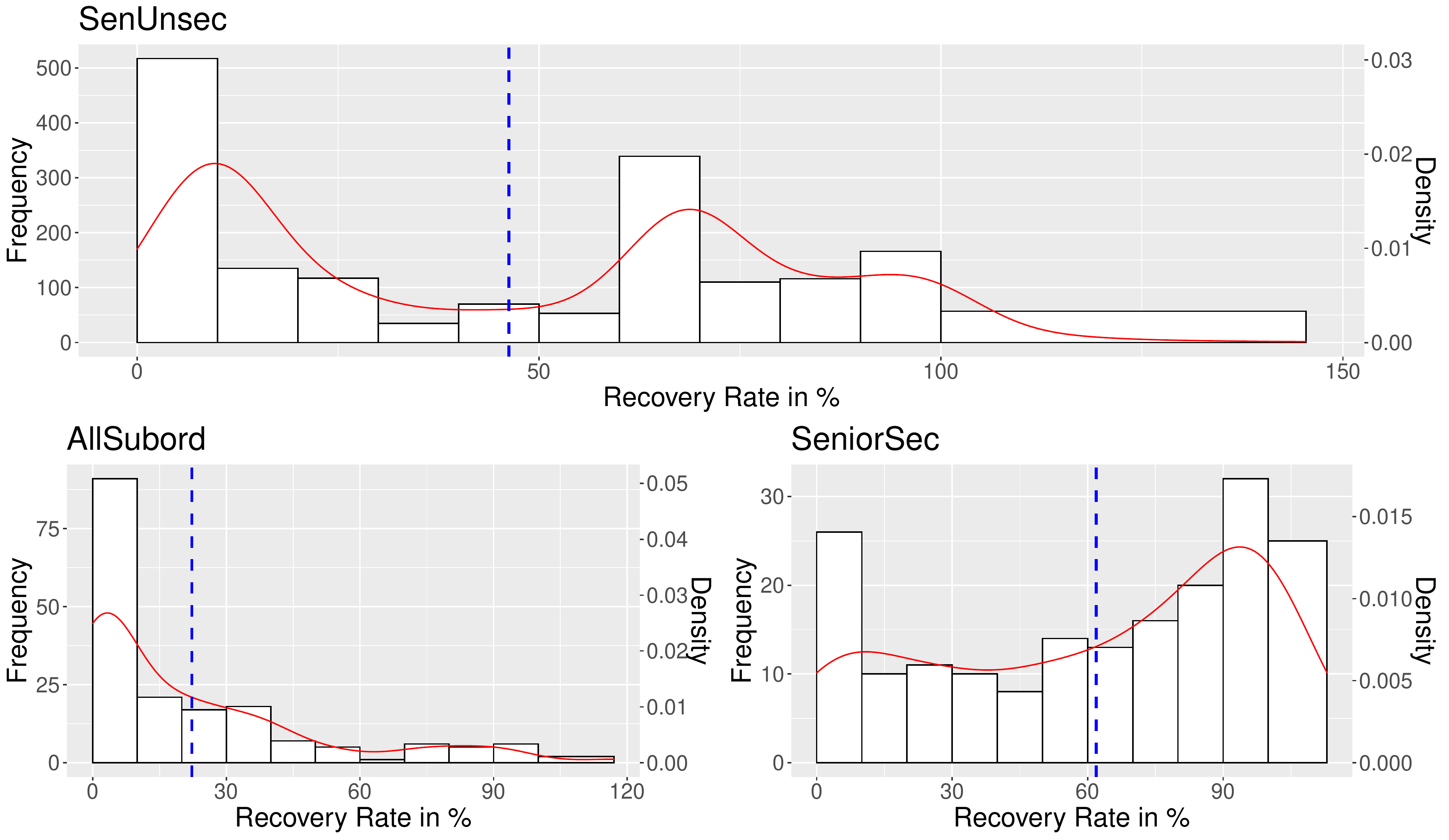}
		\caption{Recovery rate frequency and density (red) for the defaulted US corporate bonds from 2001 to 2016 and mean recovery rate in dashed line (blue). Defaults are sorted by bond type, i.e. senior unsecured bonds (\textit{SenUnsec}, $n=1715$), all subordinate bonds (\textit{AllSubord}, pooled because of insufficient data, $n=178$, from subordinate bonds: $n=158$ and senior subordinate bonds, $n=21$), and senior secured bonds (\textit{SenSec}). Please also notice the different scaling of the x-axis.}
		\label{fig:RR_bond_classes}
	\end{figure}
	
	\begin{figure}[htb]
	    \centering
	    \begin{subfigure}[b]{0.49\textwidth}
	        \includegraphics[width=\textwidth]{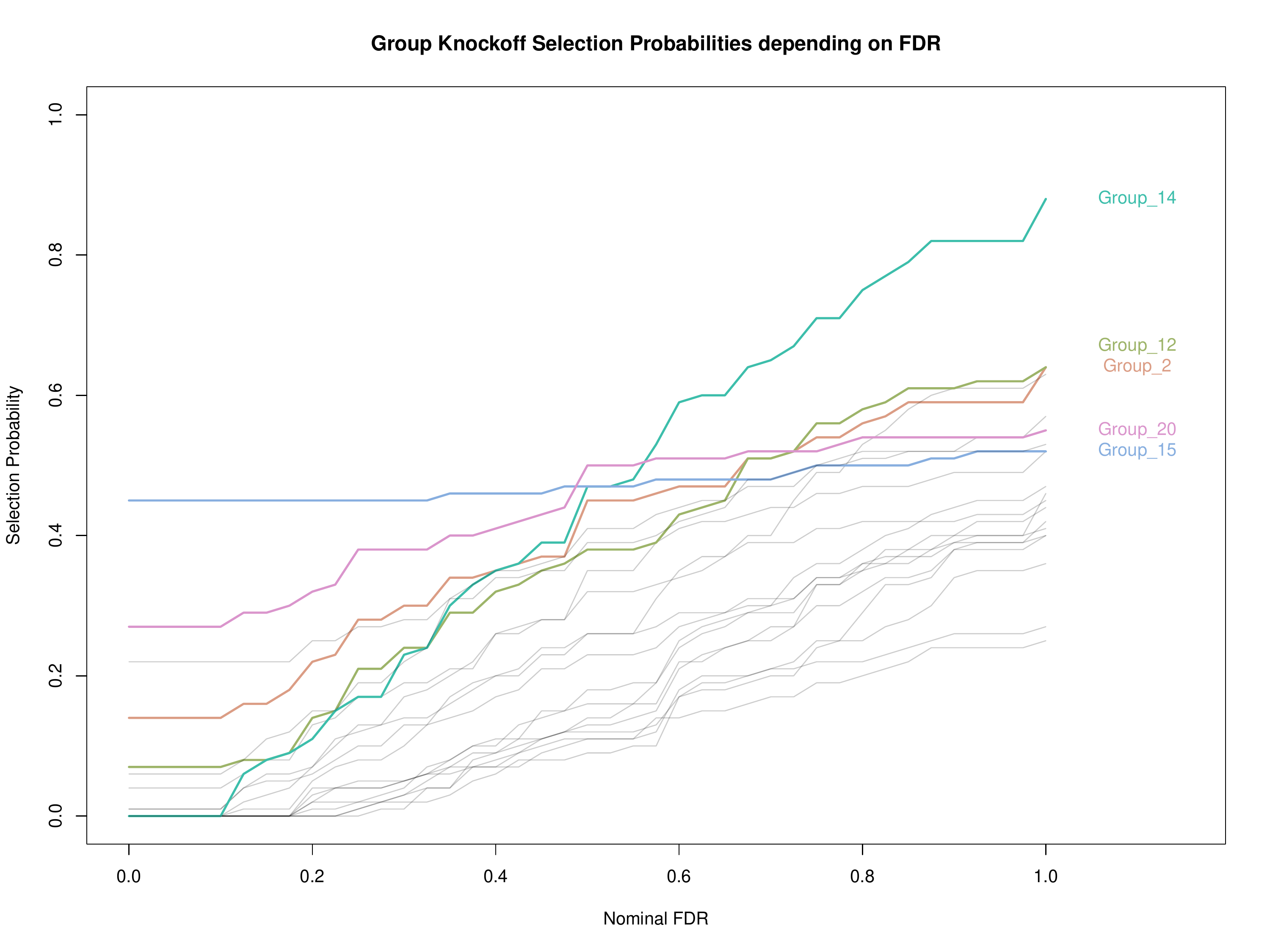}   
	    \end{subfigure}
		\hfill
		\begin{subfigure}[b]{0.49\textwidth}
		    \includegraphics[width=\textwidth]{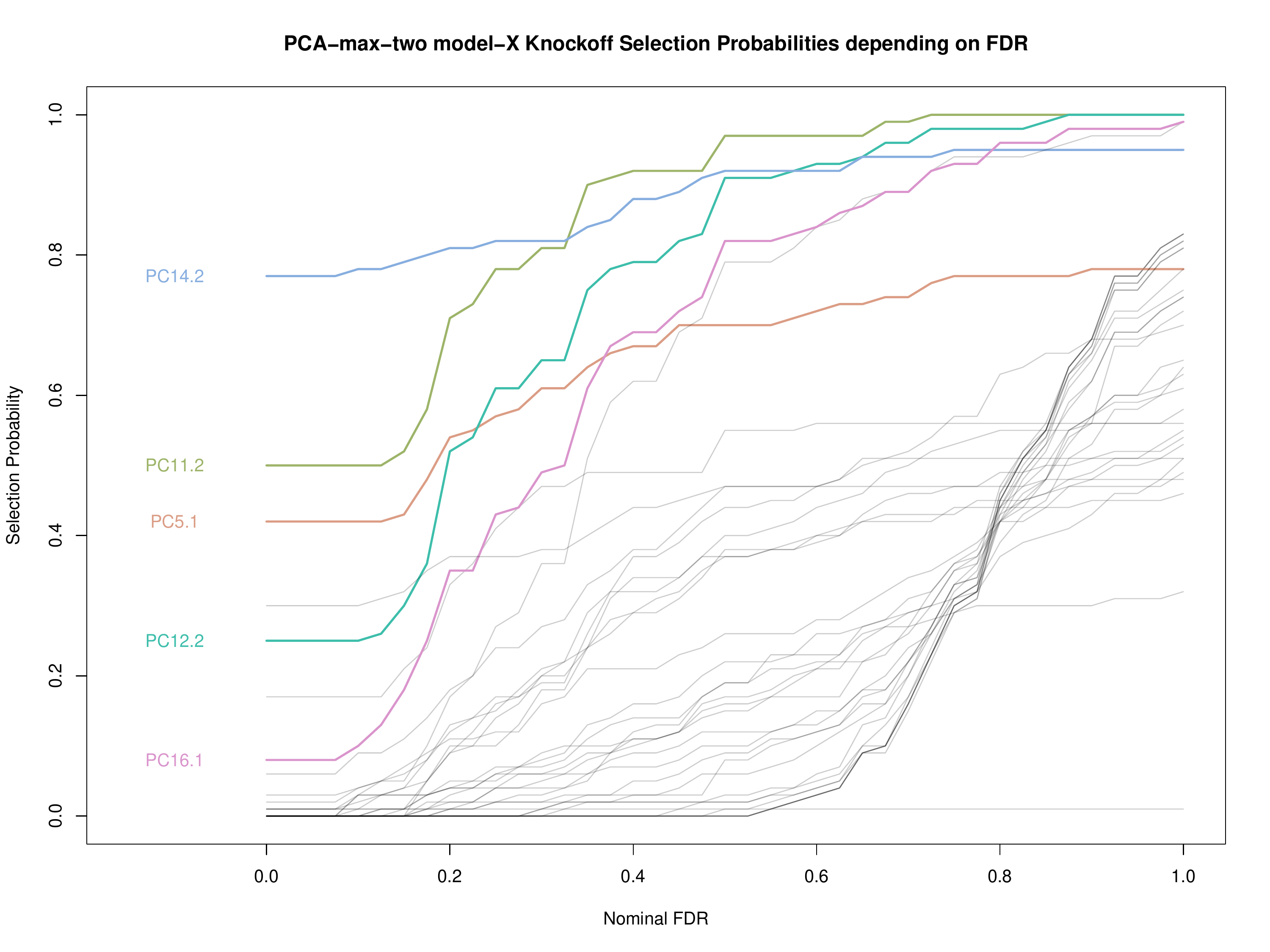}
		\end{subfigure}
		\hfill
	    \begin{subfigure}[b]{0.49\textwidth}
	       \includegraphics[width=\textwidth]{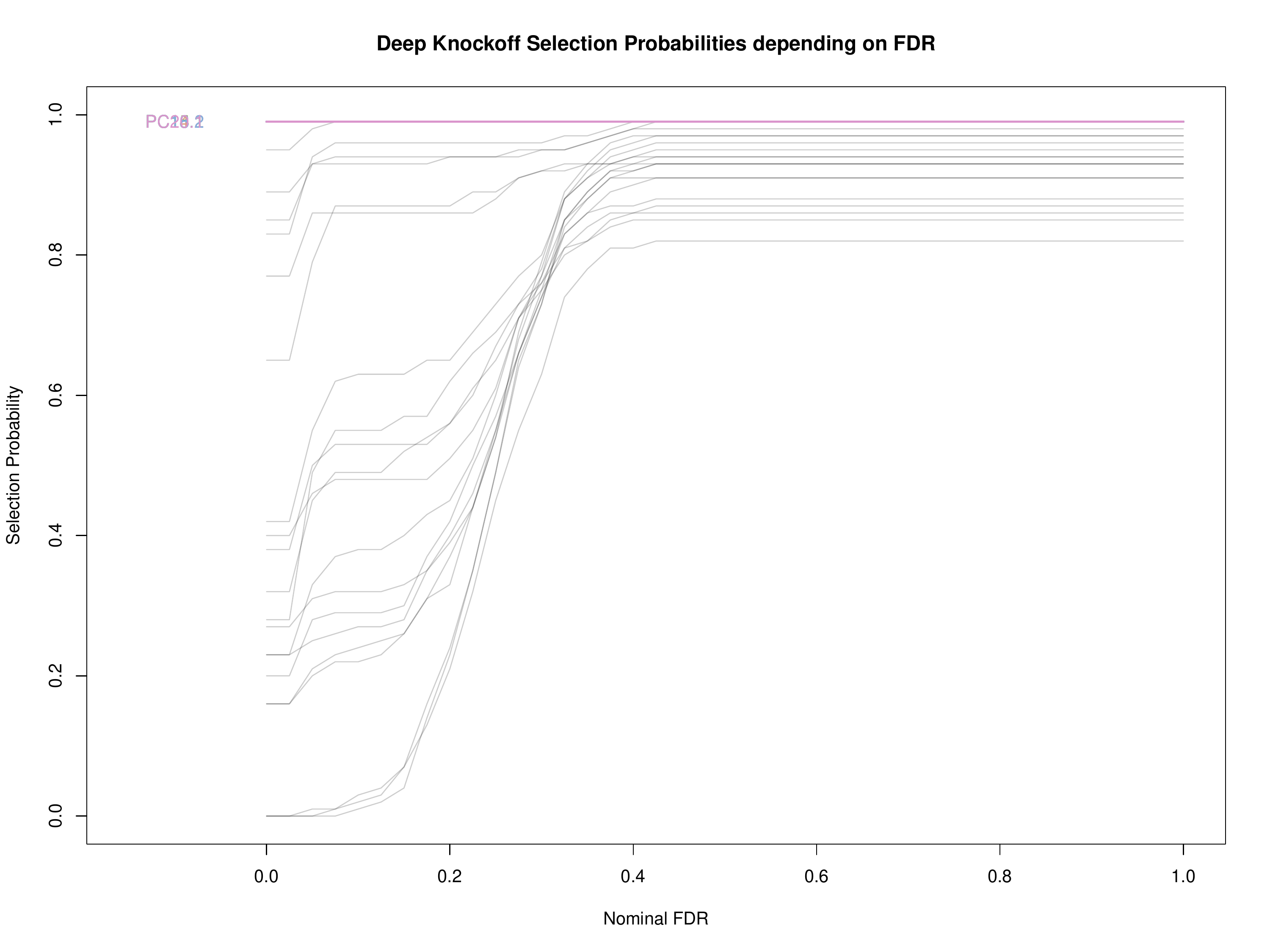}
	    \end{subfigure}
		\hfill
		\begin{subfigure}[b]{0.49\textwidth}
		   \includegraphics[width=\textwidth]{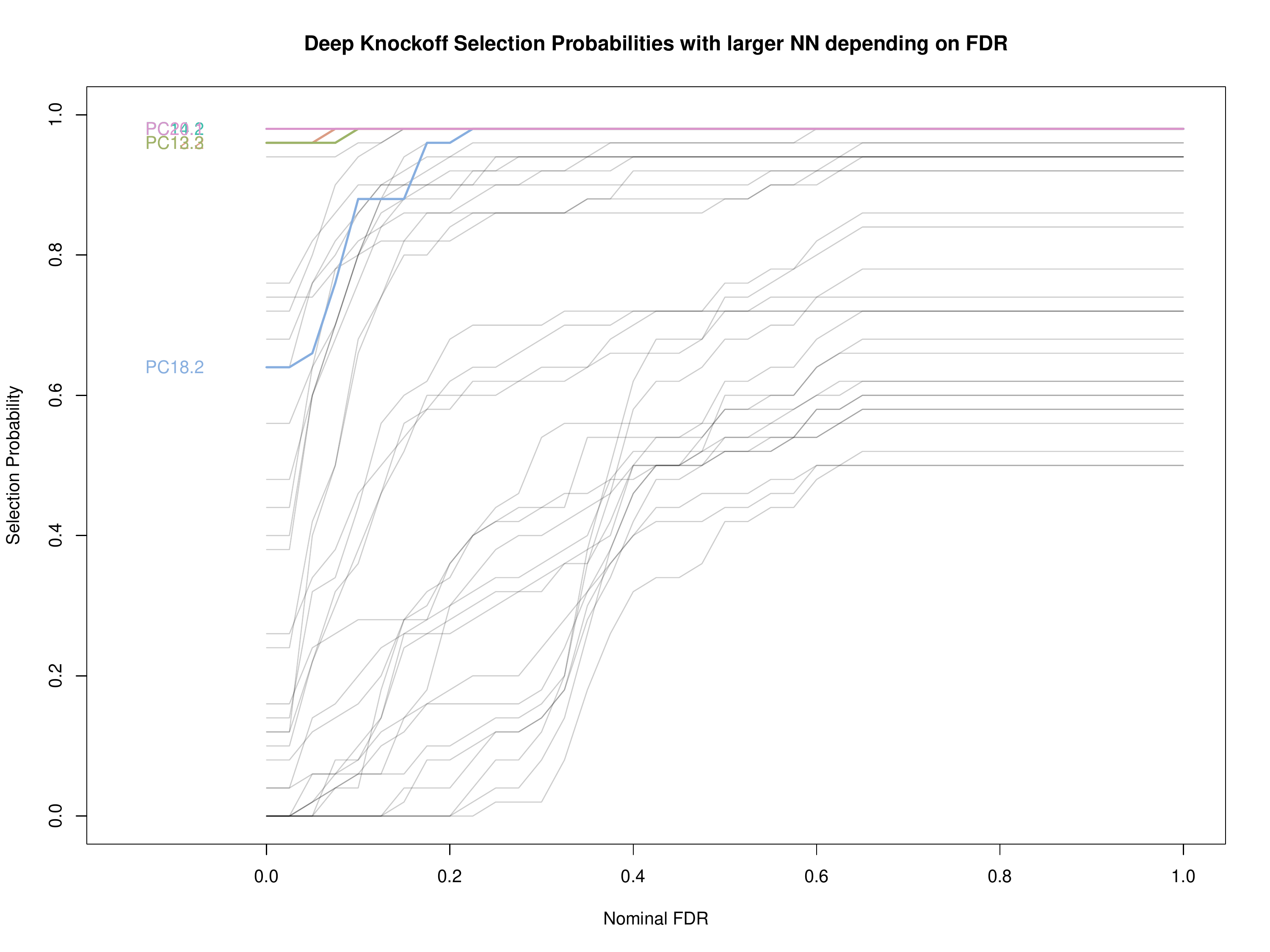}
		\end{subfigure}
		\caption{Selection Probabilities for the Group Knockoff (top-left), model-$X$ Knockoff (top-right) with maximum of two PCs per group, and Deep Knockoffs using 5 neurons (bottom-left) or 25 neurons (bottom-right) per layer. Selection Probabilities are obtained rerunning the full knockoff procedures using repeated subsampling of 90\% of the data (100 iterations). Highlighted groups have the highest mean selection rank, i.e. the mean over the rank in each FDR-scenario. The group with the highest probability receives the highest rank ($=41$ or $=20$ for the Group knockoffs) and vice versa ($=1$).}
		\label{fig:selection_probs_add_FDR}
	\end{figure}

\end{document}